%% file: ms.tex
\documentclass{emulateapj}
\newcommand\pelr{$P$-$L$}

\begin{document}
\submitted{{\sc Accepted to AJ:} March 11, 2009}
\title{The ACS Nearby Galaxy Survey Treasury III: Cepheids in the Outer Disk of M81}
\shortauthors{McCommas et al.}
\shorttitle{ANGST III:  Cepheids in M81}

\author{Les P. McCommas\altaffilmark{1}, Peter Yoachim\altaffilmark{2},  Benjamin F. Williams\altaffilmark{1}, Julianne J. Dalcanton\altaffilmark{1},Matthew R. Davis\altaffilmark{1}, Andrew E. Dolphin\altaffilmark{3}}

\altaffiltext{1}{Astronomy Department, University of Washington, Box 351580, Seattle, WA 98195; lmccomma@astro.washington.edu, jd@astro.washington.edu, ben@astro.washington.edu, mrdavis@astro.washington.edu}

\altaffiltext{2}{Department of Astronomy and McDonald Observatory, University of Texas, Austin, TX 78712; yoachim@astro.as.utexas.edu}

\altaffiltext{3}{Raytheon Corporation, 870 Winter Street Waltham, MA 02451; adolphin@ratheon.com}

\begin{abstract}
The ACS Nearby Galaxy Survey Treasury (ANGST) has acquired deep ACS
imaging of a field in the outer disk of the large spiral galaxy M81.
These data were obtained over a total of 20 HST orbits, providing a
baseline long enough to reliably identify Cepheid variable stars in
the field.  Fundamental mode and first overtone types have been
distinguished through comparative fits with corresponding Cepheid
light curve templates derived from principal component analysis of
confirmed Cepheids in the LMC, SMC, and Milky Way.  A distance modulus
of $27.78\pm0.05_r\pm0.14_s $ with a corresponding distance of 3.60
$\pm$ 0.23 Mpc has been calculated from a sample of 11 fundamental
mode and 2 first overtone Cepheids (assuming an LMC distance modulus
of $\mu_{LMC}=18.41\pm0.10_r\pm0.13_s$).
\end{abstract}

\section{Introduction}
The ACS Nearby Galaxy Survey Treasury (ANGST) is acquiring resolved stellar photometry with Hubble Space Telescope to determine the star formation history of our local volume of the universe \citep{Dalcanton08}.  The survey includes repeated long exposures for some targets, such as M81, to resolve the faintest stars possible.  Having many images of the same field over several epochs makes it possible to identify bright variable stars including Cepheids.    

Cepheid variable stars are used widely for extragalactic distance determinations, because of the firm correlation between their period of pulsation and their average absolute magnitude. They are therefore reliable standard candles, and crucial tools for determining the Hubble constant.  Other methods commonly used to measure extragalactic distances include the tip of the red giant branch (TRGB) which is often used to calibrate the Cepheid Period-Luminosity (PL) relation and vice versa.      

Once variable stars are identified from time-series photometry, the subset of Cepheid variables can be selected in several ways.  The most straightforward method is by visual inspection of the light curves, which show a characteristic saw-tooth shape.  Other more quantitative methods involve fitting the observed light curve to Cepheid templates \citep{Stetson96, Tanvir05}.  We use the Cepheid light-curve templates and fitting procedure presented in \citet{Yoachim09c}.  These templates were built by performing principal component analysis (PCA) on a large sample of Galactic Cepheids \citep{Berdnikov95,Moffett84} as well as LMC and SMC stars \citep{1999AcA....49..223U,1999AcA....49..437U}, similar to the procedure of \citet{Tanvir05}.

As M81 is one of the most massive spirals in the Local Volume, there is a long history of attempts to measure its distance \citep[e.g.,][]{Hubble29}.  Ground based observations have resulted in Cepheid distances with large uncertainties of $\Delta\mu\sim$0.30 \citep{Madore93}.  The Hubble Key Project greatly improved upon previous measurements and reported a distance modulus for M81 of $27.75\pm0.07$.  The Key Project distance was based on 25 long-period Cepheid light curves observed over 18 epochs in a field $\sim$1-2 disk scale lengths from the galaxy's center \citep{Freedman01,Freedman94}.  There have also been numerous recent studies using the TRGB method \citep{2007ApJ...661..815R,2005A&A...431..127T, Dalcanton08}.  The TRGB has the advantage of not requiring multiple epochs of data to calculate a distance, but is still a tertiary distance indicator relying on calibration from Cepheids or other distance measurements.

In this paper, we show that accurate distances can now be calculated using sparsely sampled short-period Cepheids without an observing campaign optimized for time sampling.  This advance is made possible by the combination of high-accuracy photometry from HST/ACS combined with template light curves.  Moreover, in this study we use an outer disk-field which should have substantially less extinction and crowding than previous studies.  In Section~\ref{data}, we describe our observations and data reduction techniques. In Sections~\ref{cephs} and~\ref{dist}, we isolate Cepheid variables and use them in distance calculations.  We include relevant tables and light curves in an Appendix.

\section{Observations and Data Reduction}\label{data}

\begin{deluxetable}{l c c c c}[h]
\tablewidth{0pt}
\tabletypesize{\scriptsize}
\tablecaption{Observation Log \label{obs_log}}
\tablehead{\colhead{File Name}& \colhead{Date} & \colhead{UT} & \colhead{Filter} & \colhead{(sec)} }
\startdata
j9ra58tpq\_flt.fits & 2006-11-17 & 18:43:34 & F606W & 2708\\
j9ra58tqq\_flt.fits & 2006-11-17 & 20:17:17 & F814W & 2735\\
j9ra59tuq\_flt.fits & 2006-11-17 & 21:59:17 & F606W & 2468\\
j9ra59tyq\_flt.fits & 2006-11-17 & 23:33:01 & F814W & 2495\\
j9ra60o9q\_flt.fits & 2006-11-16 & 18:45:42 & F606W & 2708\\
j9ra60oaq\_flt.fits & 2006-11-16 & 20:19:15 & F814W & 2735\\
j9ra61ojq\_flt.fits & 2006-11-16 & 23:33:16 & F606W & 2708\\ 
j9ra61okq\_flt.fits & 2006-11-17 & 01:06:51 & F814W & 2735\\
j9ra62wvq\_flt.fits & 2006-11-18 & 18:41:25 & F606W & 2708\\ 
j9ra62wwq\_flt.fits & 2006-11-18 & 20:15:16 & F814W & 2735\\
j9ra63wzq\_flt.fits & 2006-11-18 & 21:53:08 & F606W & 2708\\
j9ra63x0q\_flt.fits & 2006-11-18 & 23:27:00 & F814W & 2735\\
j9ra64n7q\_flt.fits & 2006-11-22 & 21:44:10 & F606W & 2708\\
j9ra64n8q\_flt.fits & 2006-11-22 & 23:18:47 & F814W & 2735\\
j9ra65dtq\_flt.fits & 2006-11-20 & 13:49:06 & F606W & 2708\\ 
j9ra65duq\_flt.fits & 2006-11-20 & 15:23:36 & F814W & 2735\\
j9ra66ebq\_flt.fits & 2006-11-20 & 21:48:23 & F606W & 2708\\
j9ra66ecq\_flt.fits & 2006-11-20 & 23:22:54 & F814W & 2735\\
j9ra67e1q\_flt.fits & 2006-11-20 & 17:00:49 & F814W & 2708\\
j9ra67e2q\_flt.fits & 2006-11-20 & 18:34:44 & F814W & 2770
\enddata
\end{deluxetable}

\begin{figure}[h]
\epsscale{1.2}
\plotone{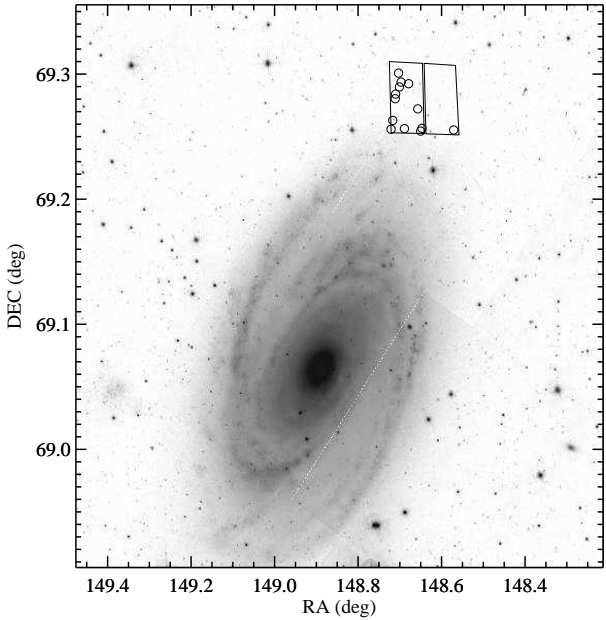} 
\footnotesize
\caption{Image of M81 from SDSS.  The ANGST M81 deep field is outlined and the location of our confirmed Cepheids are marked with circles.\label{target_field}}
\end{figure}

\begin{figure}[h]
\epsscale{1.2}
\plotone{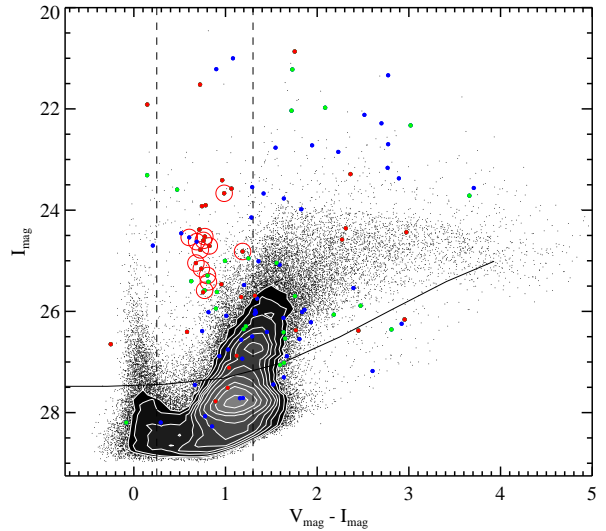}
\caption{Color magnitude diagram of M81 deep field.  All resolved stars are plotted in black.  Stars with a variability index greater than five are plotted in blue, variability index greater than nine in green, and variability index greater than fifteen in red.  Vertical dashed lines represent instability strip boundaries applied as part of Cepheid selection criteria.  Curved line represents the 50\% completeness for an individual frame.  Red circles are the final 13 confirmed Cepheid candidates.  \label{cmd}}
\end{figure}

All photometry was taken from the ANGST data products \citep{Dalcanton08, Williams08}.  Our full observation log is listed in Table~\ref{obs_log}, while the ACS target field is shown in Figure~\ref{target_field}.  ANGST data products came from the package DOLPHOT \citep{Dolphin00} which includes single frame magnitudes, combined magnitudes, data quality, and errors for each star in the field.  Standard Johnson-Cousins $V$ and $I$ magnitudes were produced by DOLPHOT which were transformed from F606W and F814W passbands using \citet{2005PASP..117.1049S}.  Catalogs were limited to high quality stellar photometry based on combined $V$ and $I$ signal-to-noise, sharpness, and crowding.  Only those stars that were resolved in every individual frame were carried through for variability index determination.  The 50\% completeness limit of an individual image was 27.5 mag for $I$ and 28.6 mag for $V$ and is indicated on the color magnitude diagram for the co-added data in Fig~\ref{cmd}.

The magnitude errors returned by DOLPHOT are extremely small.  They accurately reflect photon counting errors, but do not include systematic errors due to blending.  To assess the empirical errors, we made use of artificial star tests \citep{Williams08}.  Millions of artificial stars with known input magnitudes were inserted into each of the ACS science images.  The images were then reprocessed through DOLPHOT.  The same quality cuts of S/N, sharpness, and crowding were applied as for the real stars.  The cataloged uncertainty for each star has been updated with the standard deviation of the difference between comparable artificial stars' input and output magnitudes.  These ``comparable'' artificial stars were selected within a 100x100 pixel region centered on the real star and within a $\pm$0.2 magnitude range. The new errors generated for individual frame magnitudes using this method are slightly larger than the original magnitudes directly produced by DOLPHOT.

\begin{figure}
\epsscale{1.2}
\plotone{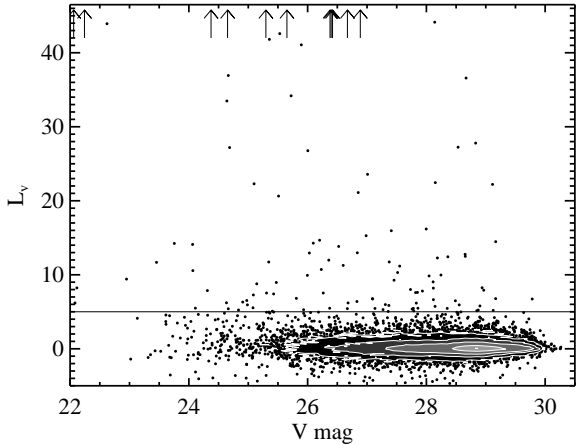}
\caption{Welch-Stetson variability index $L_v$ as a function of $V$ magnitude.  The solid line indicates the minimum $L_v$ for further Cepheid selection criteria applications.  Arrows represent stars with a variability index beyond the plotting area; the greatest of these has $L_v$ = 200. \label{var_index}}
\end{figure}

Identification of likely variables was performed by calculating a \citet{1993AJ....105.1813W} variability index for each star.  Residuals from weighted averages of each star's magnitude in each filter were used to calculate an overall variability index given by, 
\begin{equation}
L_v=\sqrt{\frac{1}{n(n-1)}}\sum_{k=1}^n (\delta V_k \delta I_k)
\end{equation}
where $\delta V_k$ and $\delta I_k$ are the normalized magnitude residuals in $V$ and $I$.  Figure~\ref{var_index} shows the variability index for the $10^5$ stars detected in the field.  Stars with a high variability index are marked on the full color magnitude diagram (Figure~\ref{cmd}).  Many variables form a well defined instability strip, along with a population of likely luminous red variables.

\section{Cepheid Selection Criteria}\label{cephs}

We attempted to fit the Cepheid light curve templates detailed in \citet{Yoachim09c} to the individual frame magnitude data for the 114 stars with a variability index greater than five.  The 114 stars were run through two passes of the Cepheid light curve template fitting procedure, once attempting fits to a short period fundamental mode template and once attempting fits to a first overtone template.  

Cepheid light-curve templates were derived from principal component analysis of Galactic Cepheids \citep{Berdnikov95,Moffett84} as well as LMC and SMC stars \citep{1999AcA....49..223U,1999AcA....49..437U} similar to the procedure used in \citet{Tanvir05}.  Unlike previous studies, \citet{Yoachim09c} generate templates for short period ($<$10 days) and overtone Cepheids.  Normally, it would take approximately 20 parameters to accurately fit a well sampled variable star light curve using a Fourier decomposition.  By using PCA, we can reduce the dimensionality of the problem and generate accurate light curves with only four free parameters ($I$ and $V$ magnitudes, period, and phase).  

We combine the PCA light curve templates with a Levenberg-Marquardt least-squares fitting routine\footnotemark[1] that returns the best fitting periods and magnitudes along with uncertainties.    
\footnotetext[1]{The Marquardt least-squares fitting routine can be found at http://cow.physics.wisc.edu/$\sim$craigm/idl/fitting.html}
Light-curve data in both filters from all of the flagged variables were run though the least-squares procedure with the Cepheid templates.  We ran the fitting procedure with a variety of initial guess parameters to ensure we converged on the global  $\chi^2$ minimum.

\begin{deluxetable}{l c c}
\tablewidth{0pt}
\tablecaption{ Effects of Selection Criteria}
\tablehead{\colhead{Selection Criteria } & \colhead{ FU } &  \colhead{FO} }
\startdata
 Initial Number & 114 & 114 \\
 1. $\chi^2$/dof $<$ 7.................... & 64 & 54 \\
 2. $0.25<(V-I)<1.3$..... & 34 & 22 \\
 3. $1<P<10$.................... & 25 & n/a \\
 4. $0.4<P<6.3$................ & n/a & 15 \\
\enddata
\end{deluxetable}

The best Cepheid light curve fits were determined for each of the 114 stars with an $L_v > 5$.  For fundamental mode template fits the number of candidate Cepheids was reduced to 64 by accepting only high quality fits with a maximum $\chi^2/DOF$ of 7.  Cuts based on position on the color magnitude diagram were also applied to further segregate true Cepheids.  A conservative color boundary of $0.25 < (V-I) <$ 1.3 ensured that only those stars that lie within the instability strip were included.  This further reduced our number of possible fundamental Cepheids to 34.  Of these, only variables with a period range of 1-10 days were retained.  Tests show that our template fitting procedure is only accurate if a substantial portion of the full Cepheid phase is observed.  Because our observations span only a 6.19 day baseline, we reject any fit that converges on 10 or more days as unreliable reducing the number of candidates to 25.  We clearly detect several long period Cepheids in the field, but can only constrain their periods to within a few days, making them unsuitable for distance determinations.  Similar selection criteria were applied to the first overtone template fits with the only difference being the requirement that the period fall within $0.4 < \rm{Period} <6.3$ days.  This acceptance range was based on data on first overtone Cepheids in the LMC from the OGLE Cepheid study \citep{1999AcA....49..223U}.  After applying our full set of selection criteria, we had 29 stars remaining.  Of these remaining candidates, we found 25 could be well fit with fundamental mode templates and 15 could be well fit with overtone templates.  Results of these selection criteria are summarized in Table 2.  

\begin{figure}
\epsscale{1.2}
\plotone{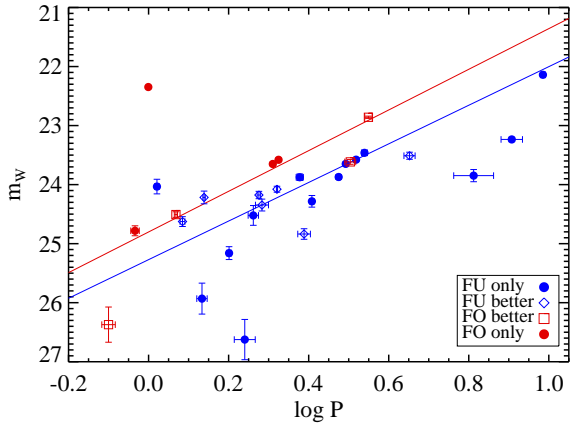}
\caption{Period-luminosity plot for candidate Cepheids in the outer disk of M81.  Blue circles are stars that passed only the fundamental mode selection criteria.  Red circles are stars that passed only the first overtone criteria.  Blue diamonds are those that passed the selection criteria for both types but have a better $\chi^2$ fit to the fundamental mode template.  Red squares fit both but have a better $\chi^2$ fit to the first overtone template.  The solid lines represent the OGLE \pelr\  relations for fundamental mode(bottom) and first overtone(top) Cepheids with the zero-point offset accounting for metallicity and the distance modulus to M81 calculated in this paper ($\mu_o=27.8\pm0.05_r\pm0.14_s$). \label{plplot}}
\end{figure}

We compute reddening-free apparent magnitudes $m_W$ \citep{1982ApJ...253..575M} for all our candidate stars and plot an initial period-luminosity relation in Figure~\ref{plplot}.  As was expected, many Cepheid candidates passed all of the selection criteria for the fundamental mode type and for both the first overtone type.  Those that passed both sets of criteria were sorted into two additional categories: those that have a better $\chi^2$ fit to the fundamental light curve template and those that have a better $\chi^2$ fit to the first overtone light curve template.  Some candidates were so close in their fits to both templates that it was not possible to distinguish the type; these seven candidates were removed from the list for being indistinguishable.

\begin{figure}[ht]
\epsscale{1.2}
\plotone{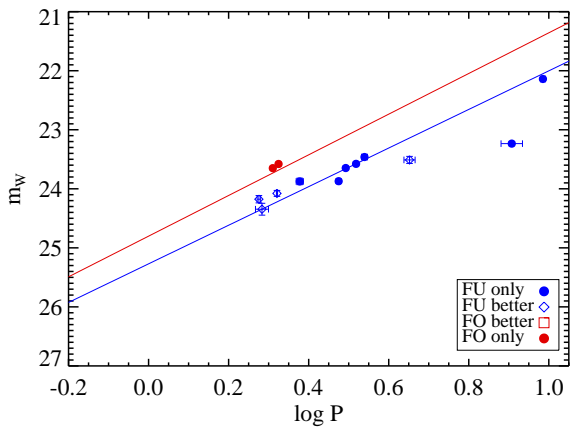}
\caption{Period-luminosity plot for the final fundamental mode and first overtone Cepheid candidates.  Symbols and solid lines are the same as in Figure 4. \label{final_pl}}
\end{figure}

Of the remaining Cepheid candidates, visual inspection of the light curves was used to make the final quality cut.  Nine additional candidates were removed in this manner.  These were typically stars where one or two outliers gave the star the appearance of variability.  With all selection criteria and quality cuts applied, 11 likely fundamental mode and 2 likely first overtone Cepheids remain.  Figure~\ref{final_pl} shows the period-luminosity diagram for these stars.  Their locations within M81 are plotted in Figure~\ref{target_field}, and they clearly lie on an extension of the inner spiral arm.  These final results match well with period-luminosity relations derived from Udalski 1999 with the zero-point adjusted for metallicity and distance modulus to M81.  The final fit parameters and light curves are presented in Figures~\ref{lc1} and~\ref{lc2}.  The full light-curve photometry points are provided in Tables~6~\&~7.

\section{Period Luminosity Relations and Distance Calculations}\label{dist}

We assume the period-luminosity (\pelr) relations and errors for standard $V$ and $I$ magnitudes derived from recent OGLE Cepheid studies of the LMC \citep{1999AcA....49..223U}\footnotemark[2].  We have several reasons for using the LMC derived \pelr\  relation over other popular \pelr\  relations \citep[e.g.,][]{Freedman01, Sandage04,Benedict07}.  First, The OGLE studies include overtone \pelr\  relations.  Second, the OGLE sample is dominated by short-period stars, like those in our study, and it is not clear that \pelr\  relations derived from long period stars can be accurately extrapolated, as there is a possible discontinuity in the \pelr\  relation around a period of ten days \citep{Kanbur04}.  Finally, because our observations are in an outer field of M81, we expect a fairly good match between our Cepheid metallicities and those in the LMC, meaning we only need to make a very small metallicity adjustment to the \pelr\  relation (\S\ref{met_core}).  The major disadvantage in using the OGLE \pelr\  relation is that our distance determination is explicitly tied to the LMC distance, which becomes our dominant source of uncertainty.  The adopted \pelr\ relations are:

\footnotetext[2]{Updated OGLE Cepheid PL relations can be found at ftp://sirius.astrouw.edu.pl/ogle/ogle2/var\_stars/lmc/cep\\ /catalog/README.PL.}

\begin{center}
Fundamental Mode \pelr\  Relations
$V_{LMC} = -2.779(31)\log \rm{P} + 17.066(21)$ \\
$I_{LMC} = -2.979(21)\log \rm{P} + 16.594(14)$ 
\end{center}

\begin{center}
First Overtone \pelr\  Relations
$V_{LMC} = -3.326(54)\log \rm{P} + 16.634(20)$ \\
$I_{LMC} = -3.374(35)\log \rm{P} + 16.147(13)$
\end{center}

\subsection{Extinction Correction}

The presence of intervening dust along the line of sight causes some Cepheids to appear fainter and redder than they would in the absence of extinction, thereby making them appear to be more distant.  Dust attenuates the $V$ passband more than the $I$ passband making distances calculated using the $V$ \pelr\  relation more distant than the ones calculated using the $I$ \pelr\  relation.  This effect was observed when single passband distances were calculated in this study. 

The effect of reddening can be corrected using the ``Wesenheit reddening-free index'' \citep{1982ApJ...253..575M},
\begin{center}
$\mu_W \equiv \mu_V - A_V = \mu_I - A_I.$  
\end{center}
For $V$ and $I$ photometry the Wesenheit index is defined as $W = V - R\times(V -I)$.  $R$ is taken to be 2.45 based on \citet{1989ApJ...345..245C} and as used in \citet{Macri06}.  The Wesenheit index then becomes $W = -1.45V + 2.45I$ for purposes of error propagation.  Using this we can write new \pelr\  relations for the Wesenheit index:

\begin{center}
Fundamental Mode \\
$W_{LMC} = -3.269(68)\log \rm{P} + 15.910(46)$  
\end{center}

\begin{center}
First Overtone \\
$W_{LMC} = -3.444(110)\log \rm{P} + 15.441(43)$
\end{center}

\citet{1999AcA....49..201U} derive their own Wesenheit corrected \pelr\  relations from least squares fitting to Wesenheit magnitudes.  These could have been used directly in this paper, however, they use a slightly different value for R than we have adopted.

\subsection{Metallicity Correction}\label{met_core}

Many Cepheids seem to show a dependence of absolute brightness on metallicity \citep{Macri06,Romaniello08, Saha06,Sandage08}, such that metal-rich Cepheids are brighter than metal-poor Cepheids of the same pulsation period.  We correct the distance moduli for this effect as follows.  \citet{1994ApJ...420...87Z} measured a metallicity gradient of -0.12$\pm$0.05 dex/$h_R$ for M81 with a value of [O/H] = 9.10$\pm$0.11 at r=0.8$h_R$.  Assuming our field is located at $R\sim 5 h_R$, the metallicity for the ANGST M81 deep field is [O/H]=12+log(O/H) $\sim$8.6.  This is consistent with the metallicities derived from AGB bump and red clump from the same field \citep{Williams08}.  Using the metallicity correction of \citet{Macri06}, our Cepheid distance moduli are corrected by
\begin{center}
$\Delta\mu^{o}_{\rm{met}}=(0.29 \pm 0.09_r \pm 0.05_s)([O/H]_{\rm{M81}}-8.50) = 0.029\pm0.009_r\pm0.005_s$
\end{center}
where the ``r'' and ``s'' subscripts denote random and systematic uncertainties respectively.

\subsection{Distance Modulus to the LMC}

The period luminosity relations shown so far are all relative to the Large Magellanic Cloud.  In order to adjust the zero point such that a true absolute magnitude for our Cepheids can be calculated, the distance modulus to the LMC must be adopted and subtracted.  

We adopt the water maser relative distance between NGC 4258 and the LMC. The discovery of water masers in the active nucleus of NGC 4258 provides a very accurate distance to that galaxy.  Using the orbits of these masers, \citet{1999Natur.400..539H} found a geometric distance to NGC 4258 of $\mu=29.29\pm0.09_r\pm0.12_s$.  \citet{Macri06} subsequently observed 281 Cepheids in NGC 4258 and using the OGLE PL relations above, found a relative distance modulus from NGC 4258 to the LMC $\Delta\mu^o=10.88\pm0.04_r\pm0.05_s$.  Combining these results gives a distance  modulus to the LMC of $\mu^o=18.41\pm0.10_r\pm0.13_s$ \citep{Macri06}, which we adopt here. 
 
\subsection{Consistency with Previous Cepheid Observations in M81}

\begin{figure}
\epsscale{1.2}
\plotone{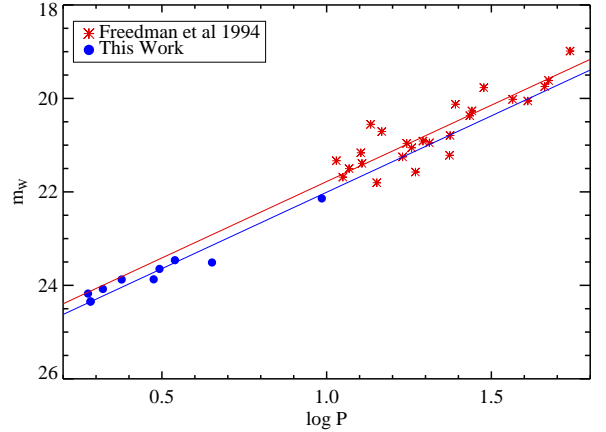}
\caption{  Our fundamental mode Cepheids (blue) compared to the observations of \citet{Freedman94} (red).  The solid lines show the best fit \pelr\ relations when the slopes are fixed at the OGLE LMC value.  There is a 0.23 magnitude offset, however this becomes statistically insignificant if we correct for the expected metallicity differences of the samples.  \label{compare}}
\end{figure}

\citet{Freedman94} also present HST observations of Cepheids in M81.  In Figure~\ref{compare} we plot the \citet{Freedman94} Wesenheit magnitudes along with our fundamental mode Cepheids.  When we fit the  \pelr\ relations, holding the slope constant at the OGLE LMC value, we find our sample has a zero-point of m$_{W_0}=25.28\pm0.05$ while the long-period Cepheids have a zero point of $25.05\pm0.07$.  The \citet{Freedman94} observations were of an inner region where we would expect the metallicity of the stars to be much higher, and therefore the inner Cepheids should appear brighter.  The metallicity gradient measured in \citet{1994ApJ...420...87Z} suggests a metallicity difference between our field and the  fields of $\sim0.5$ dex, corresponding to an expected zero-point offset of 0.14 mag.  Thus we find that the metallicity corrected zero-point offset between the two observations is $\Delta_{zp}\sim0.09\pm0.09$, consistent with no offset at the 1-$\sigma$ level.  We point out that this metallicity correction is based on an extrapolation from a slope observed in inner region HII regions.  It is conceivable that the Cepheids we observe at large radius are even more metal poor than our extrapolation guess, which would bring the \pelr\ zero-points into even better agreement.

\subsection{Distance Calculation}
Of the 11 fundamental mode and 2 first overtone Cepheids that have passed all selection criteria, individual distance moduli were calculated using the Wesenheit corrected PL relations with adjustments for metallicity ($\Delta\mu^o=0.029\pm0.009_r\pm0.005s$) and distance modulus to the LMC ($\Delta\mu^o=18.41\pm0.10_r\pm0.13_s$).  We list all our known sources of error and how they propagate to the final derived distance modulus for an individual star in Table~\ref{etable}.  We propagate random errors using standard techniques for Gaussian errors.  Our systematic errors are dominated by the systematic uncertainty in the distance to the LMC.  The weighted average of our 13 calculated distance moduli is $27.86 \pm 0.05_r \pm 0.14_s$ giving a  corresponding distance of 3.73 $\pm$ 0.24 Mpc.  

\begin{figure}[h]
\begin{center}
\epsscale{1.1}
\plotone{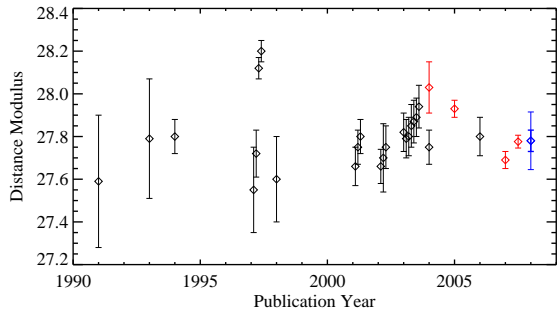}
\end{center}
\caption{Published distance moduli as a function of publication date.  Cepheid derived distances are plotted with black error bars, TRGB distances are in red, and the distance derived in this paper is in blue.  Distances compiled by \citet{Madore07}, \citet{2005A&A...431..127T}, \citet{2007ApJ...661..815R}, and \citet{Dalcanton08}.  \label{lit_dist} }
\end{figure}

Of the final group of identified Cepheids, 12 out of 13 lie within one standard deviation of the mean distance modulus.  Only one fundamental mode Cepheid lies greater than 2 standard deviations from the mean (Figure~\ref{final_pl}).  Clipping this outlier, the final distance calculation results in a distance modulus of $27.78 \pm 0.05_r \pm 0.14_s$ and a corresponding distance of $3.60 \pm 0.23$ Mpc.  It is also worth noting that the clipped star (candidate 664.580) has by far the largest error in period resulting from the PCA template fit.  This distance agrees well with previous measurements as shown in Figure~\ref{lit_dist}.  The agreement with the HST Key Project distance is excellent, in spite of the fact that we use fewer stars and non-optimally sampled light curves.  This agreement and comparable accuracy is due to the improved statistical power of using PCA light curve templates.  


\begin{deluxetable*}{l c c c}
\tablecaption{ Error Budget\label{etable}}
\tablehead{\colhead{Source } & \colhead{ Random Error  }& \colhead{ Random Error  } &   \colhead{Systematic error}\\
   & &\colhead{$\Delta \mu$} & \colhead{$\Delta \mu$} }
\startdata
Fitted Periods  &0.1-0.5 days &  0.01-0.09& \nodata \\
Fitted Magnitudes  & 0.01-0.02 mags &  0.02-0.05  & \nodata\\
$P$-$L$ relation slope & $\sim2\%$ & 0.04 & \nodata \\
$P$-$L$ relation zero point & $\sim3\%$&  0.05& \nodata \\
Reddening correction  &  & 0 & 0.06  \\
Metallicity Correction  & & 0.009& 0.005\\
$\mu_{\rm{LMC}}$  & & 0.10 & 0.13 \\
Overtone-Fundamental Classification & & \nodata & \nodata
\enddata
\end{deluxetable*}

\section{Summary}

  1.  We have isolated 11 fundamental mode and 2 first overtone Cepheid variables in an M81 deep field consisting of 9 V-band and 11 I-band images.

  2.  We calculate a distance modulus for M81 of 27.78 $\pm 0.05_r \pm 0.14_s$\ with a corresponding distance of $3.60 \pm 0.23$\ Mpc, after removing one of the confirmed Cepheid variables, due to its obvious deviation from the mean distance modulus.


  3.   The distance modulus derived in this paper is consistent with those derived in previous years (Figure~\ref{compare}).  The largest source of error in the final distance calculation is due to the systematic uncertainty in the distance modulus to the LMC ($\mu_{LMC}=18.41 \pm 0.10_r \pm 0.13s$) from \citet{Macri06}.  The combination of using the appropriate fundamental mode (P $<10$ days) and first overtone Cepheid templates of \citet{Yoachim09c} and the precision of ACS instrumentation and photometry produces uncertainties that are comparable to previous determinations of M81's distance using more stars, brighter stars, and a greater number of observed epochs.


\acknowledgments
We thank the referee Barry Madore for helpful comments that improved the paper.  Support for this work was provided by NASA through grant G0-10915 and AR-10945 from the Space Telescope Institute, which is operated by the Association of Universities for Research in Astronomy, incorporated under NASA contract NAS5-26555.


\clearpage


\appendix

\begin{deluxetable}{l c c c c c c c}
\tablewidth{0pt}
\tabletypesize{\scriptsize}
\tablecaption{ Fundamental Mode Cepheids \label{fit_param}}
\tablehead{ \colhead{ID} & \colhead{R.A. (2000)} & \colhead{Dec. (2000)} & \colhead{Period(days)} & \colhead{$<\rm{m}_V>$} & \colhead{$<\rm{m}_I>$} &  \colhead{$\mu_w$} & \colhead{$\chi^2$/dof} }
\startdata
ANGST C1  & 9:54:17.0 & 69:15:19.2 & 3.11(0.02) &   25.42(0.01)&   24.70(0.01) & 27.79(0.17) & 4.1 \\    
ANGST C2  & 9:54:52.1 & 69:15:46.7 & 9.67(0.14) &   24.53(0.01)&   23.55(0.01) & 27.89(0.17) & 5.8  \\    
ANGST C3  & 9:54:45.3 & 69:15:23.2 & 3.46(0.06) &   25.13(0.02)&   24.45(0.01) & 27.75(0.17) & 6.1 \\    
ANGST C4  & 9:54:48.3 & 69:17:22.8 & 3.30(0.03) &   25.45(0.01)&   24.69(0.01) & 27.80(0.17) & 2.0 \\    
ANGST C5  & 9:54:48.7 & 69:18:03.1 & 8.08(0.50) &   25.82(0.02)&   24.77(0.01) & 28.73(0.17) & 5.4 \\     
ANGST C6  & 9:54:35.3 & 69:15:23.8 & 2.98(0.03) &   25.57(0.01)&   24.88(0.01) & 27.95(0.17) & 2.3 \\    
ANGST C7  & 9:54:53.2 & 69:15:21.2 & 2.39(0.05) &   25.69(0.02)&   24.95(0.02) & 27.64(0.18) & 2.0 \\    
ANGST C8  & 9:54:50.6 & 69:16:49.2 & 4.49(0.14) &   25.23(0.03)&   24.53(0.02) & 28.17(0.18) & 1.6 \\    
ANGST C9  & 9:54:43.1 & 69:17:32.3 & 1.89(0.02) &   26.00(0.02)&   25.25(0.02) & 27.61(0.18) & 0.6 \\    
ANGST C10 & 9:54:36.0 & 69:15:15.9 & 1.92(0.07) &   26.17(0.04)&   25.42(0.03) & 27.80(0.19) & 1.3 \\    
ANGST C11 & 9:54:50.5 & 69:17:02.4 & 2.09(0.01) &   26.04(0.02)&   25.24(0.02) & 27.60(0.17) & 1.7     
\enddata
\end{deluxetable}

\begin{deluxetable}{c c c c c c c c}
\tablewidth{0pt}
\tabletypesize{\scriptsize}
\tablecaption{  First Overtone Cepheids}
\tablehead{ \colhead{ID} & \colhead{R.A. (2000)} & \colhead{Dec.(2000)} & \colhead{Period(days)} & \colhead{$<\rm{m}_V>$} & \colhead{$<\rm{m}_I>$} &  \colhead{$\mu_w$} & \colhead{$\chi^2$/dof}  }
\startdata
 ANGST OV1 & 9:54:37.6 & 69:16:19.9  &   2.11(0.02) &   25.10(0.01) &  24.48(0.01)   &   27.70(0.17) &  4.5\\
 ANGST OV2 & 9:54:47.4 & 69:17:36.9  &   2.04(0.02)  &  25.28(0.01) &  24.62(0.01) &     27.72(0.17) &   2.1
\enddata
\end{deluxetable}


\begin{figure}[h]
 \begin{center}$
      \begin{array}{ccc}
         \includegraphics[scale=0.3]{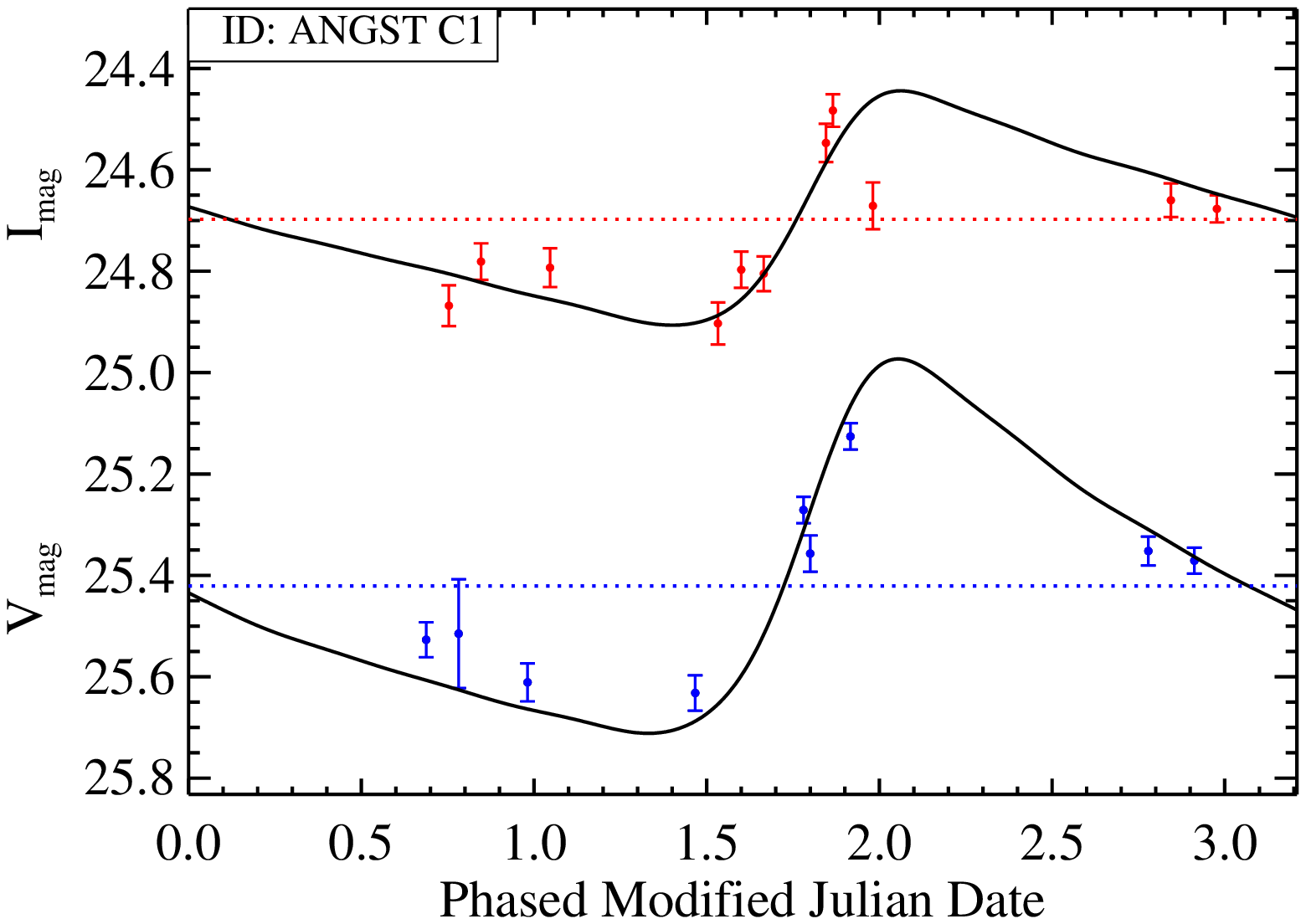} &   
         \includegraphics[scale=0.3]{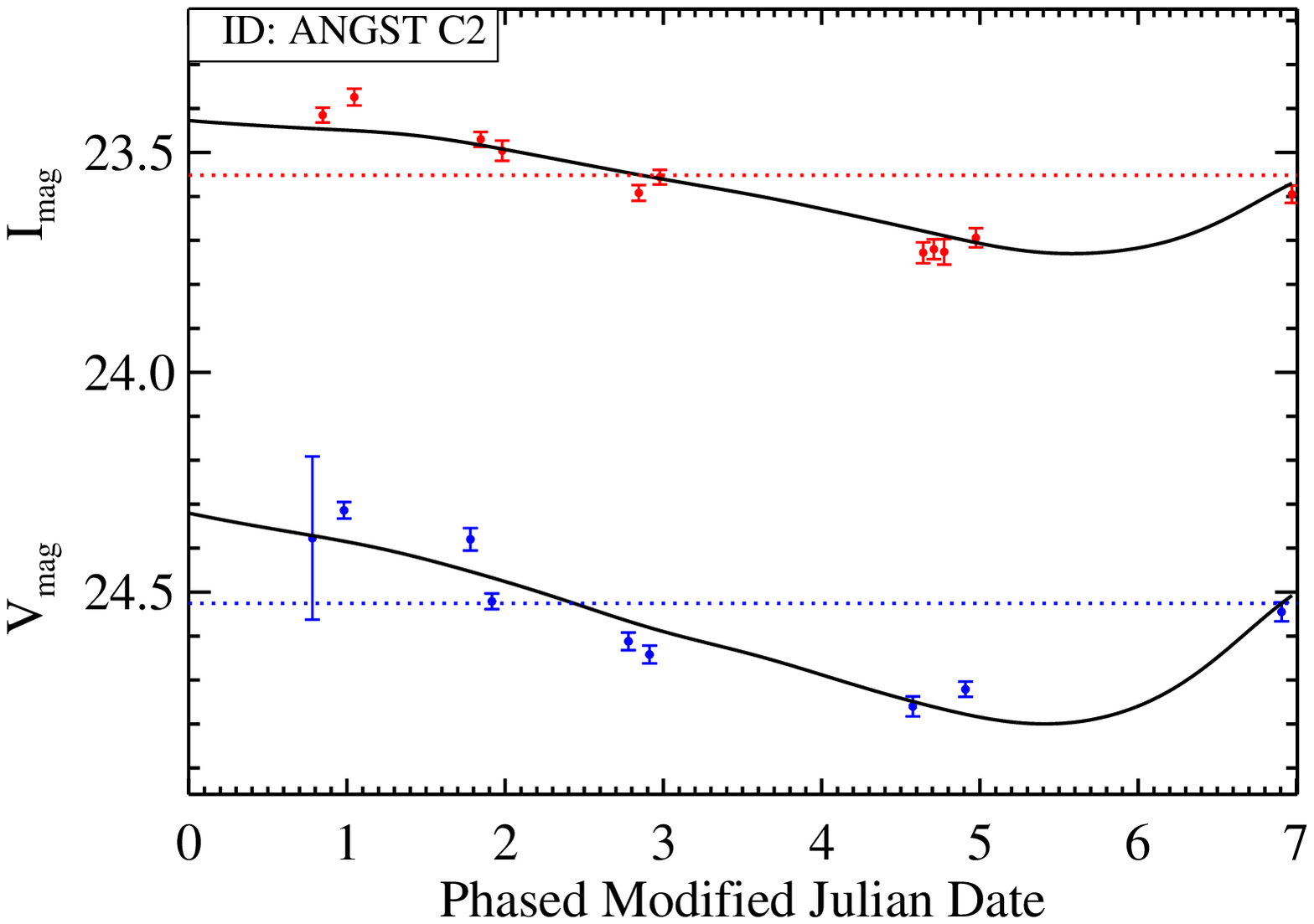} &   
	 \includegraphics[scale=0.3]{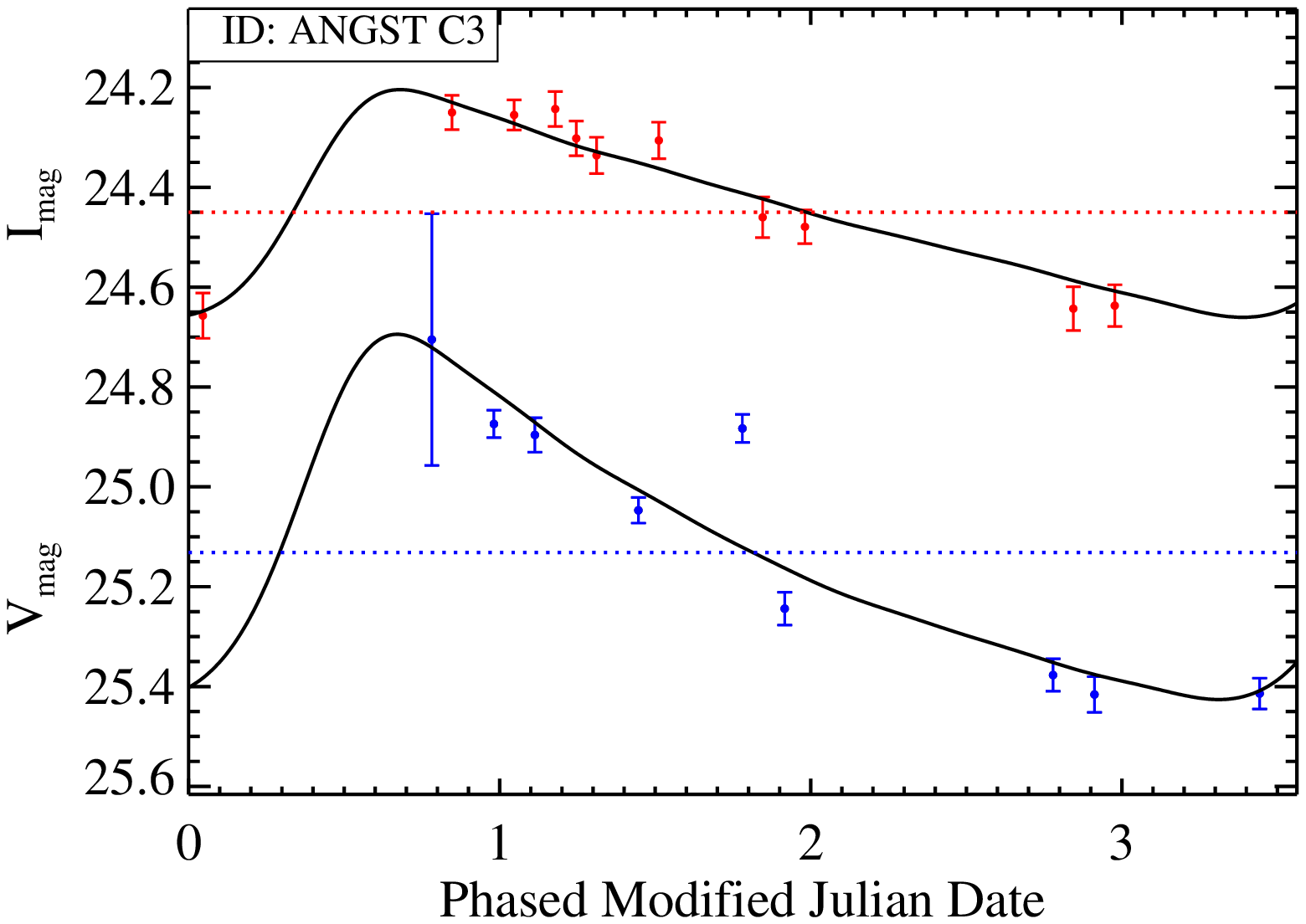} \\   
         \includegraphics[scale=0.3]{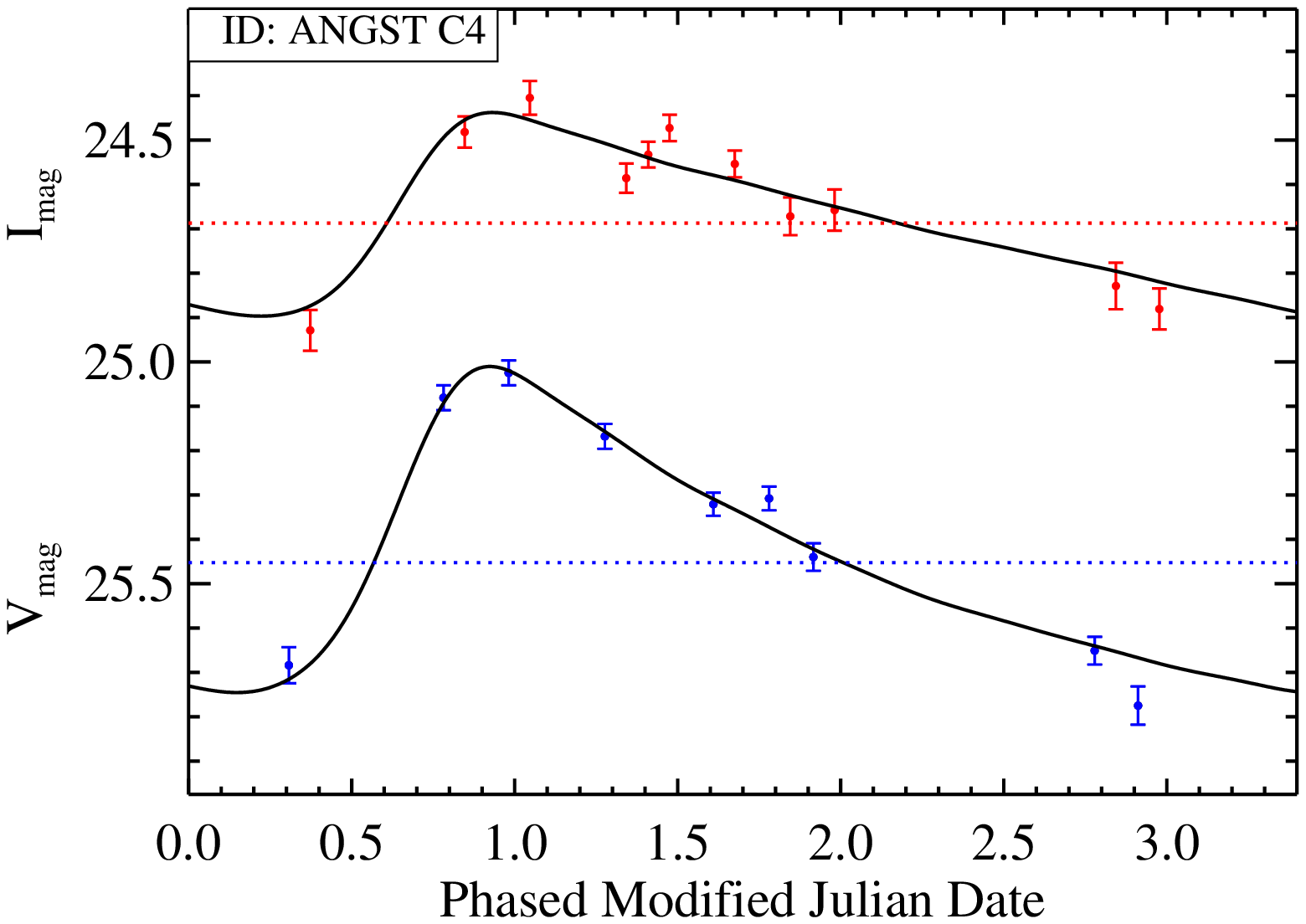} &   
         \includegraphics[scale=0.3]{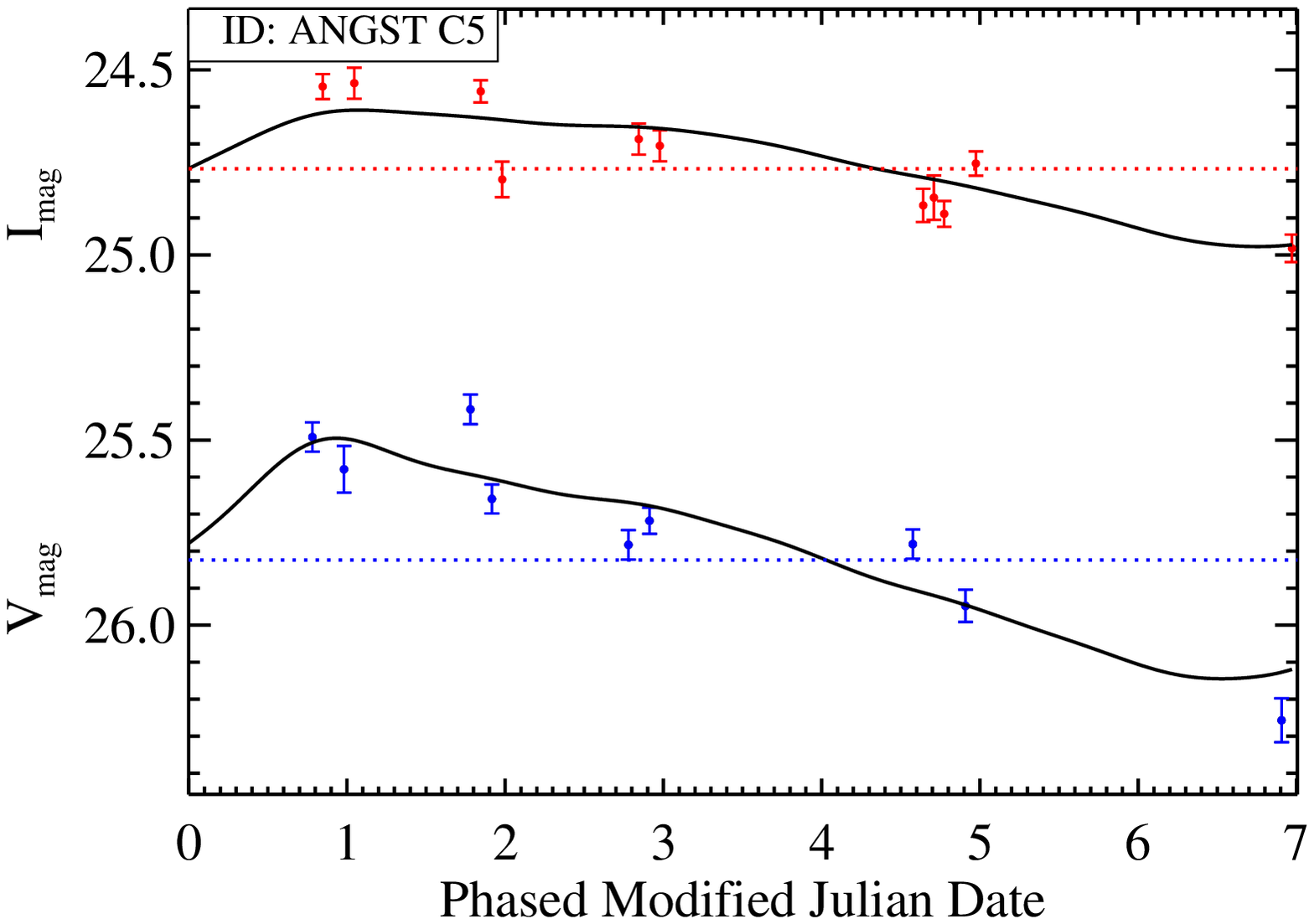} &   
         \includegraphics[scale=0.3]{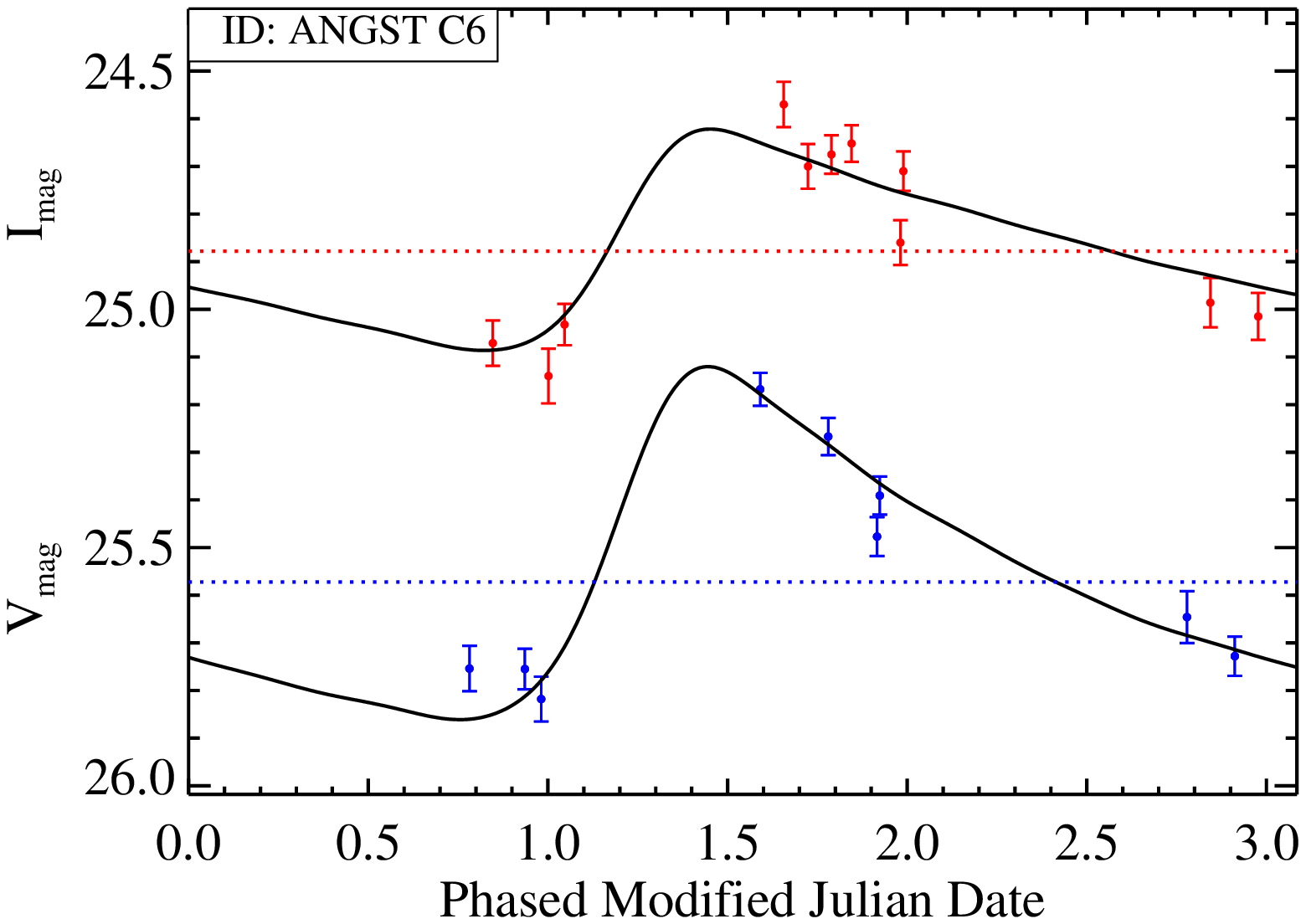} \\  
         \includegraphics[scale=0.3]{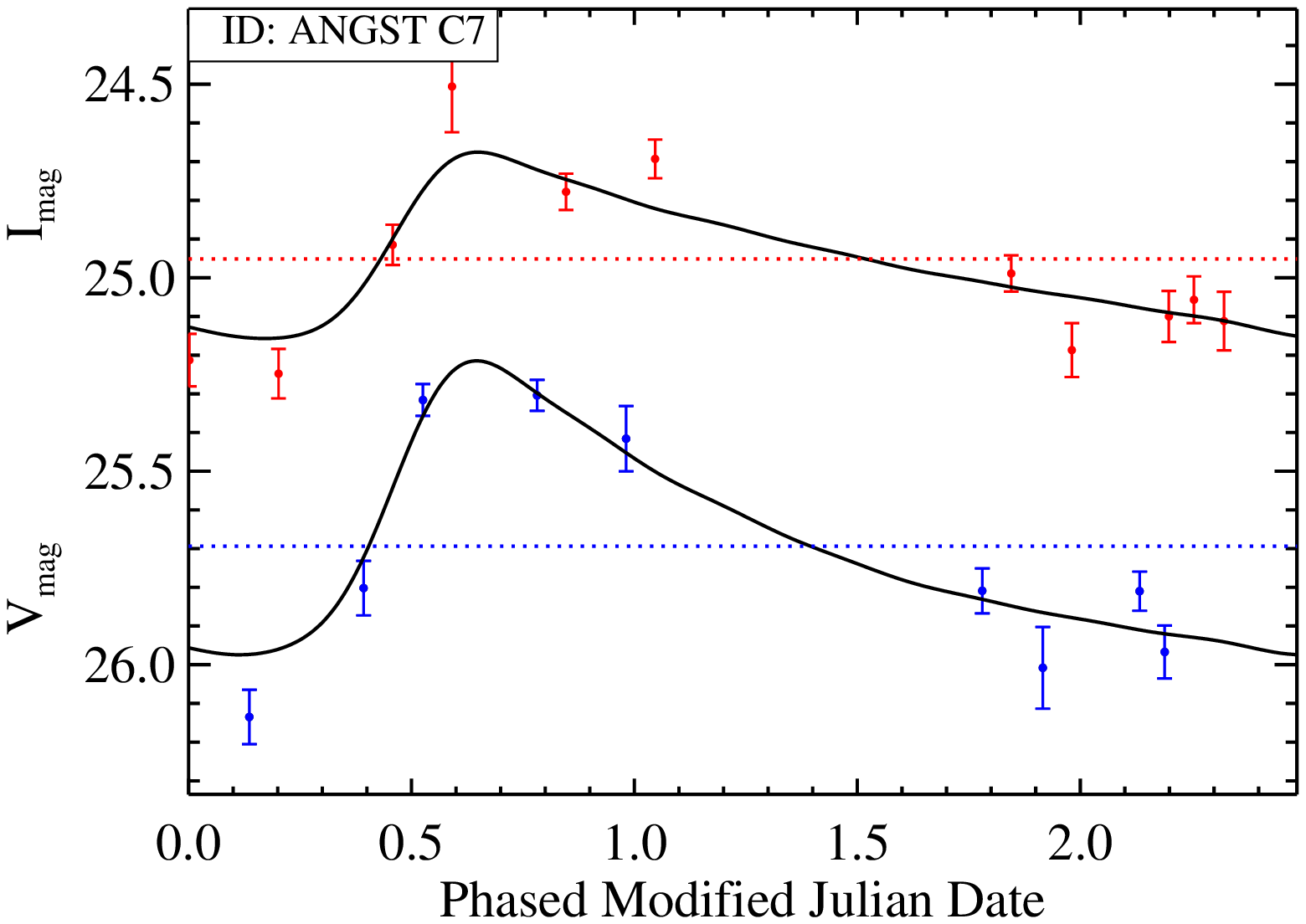} &   
         \includegraphics[scale=0.3]{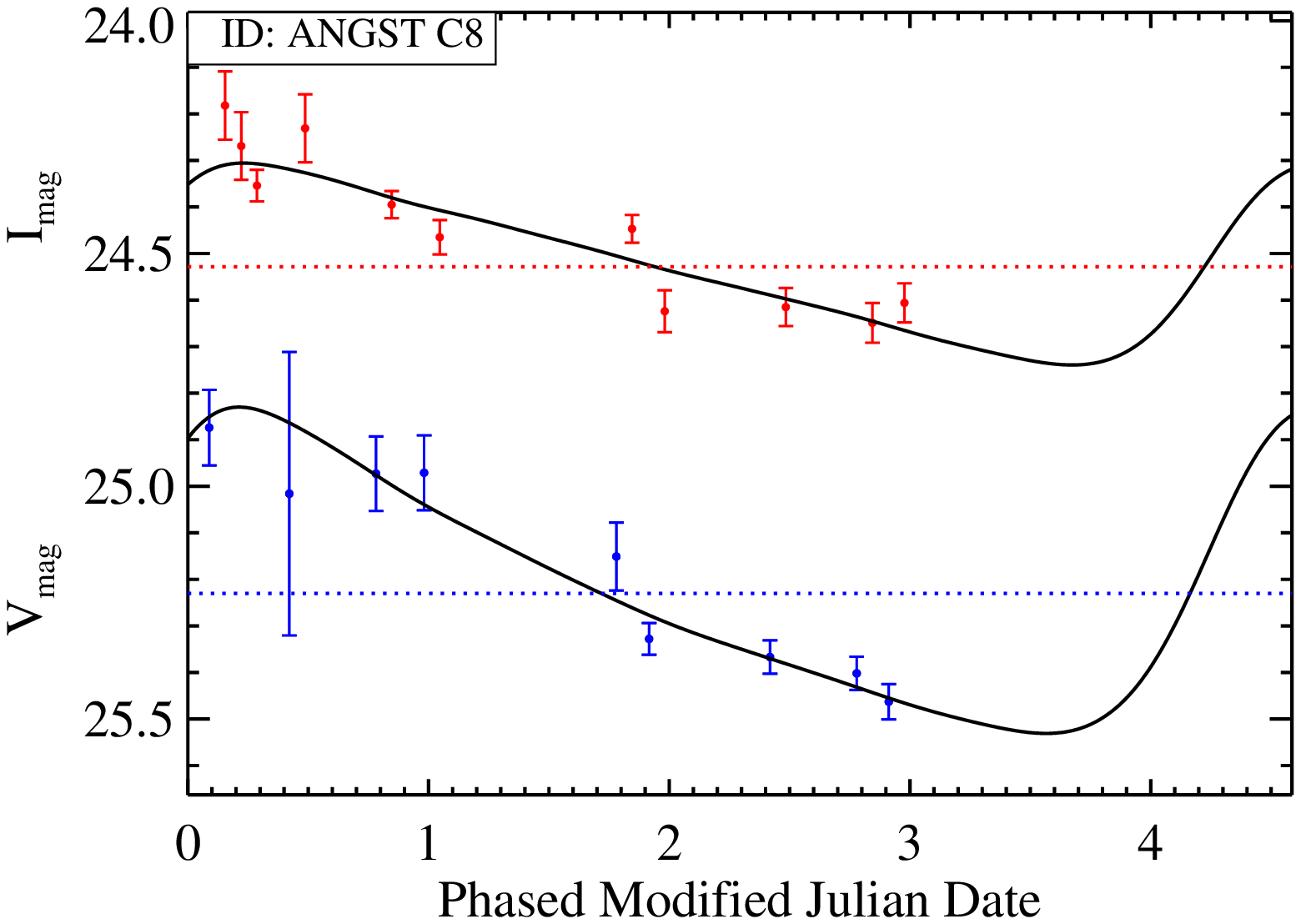} &   
         \includegraphics[scale=0.3]{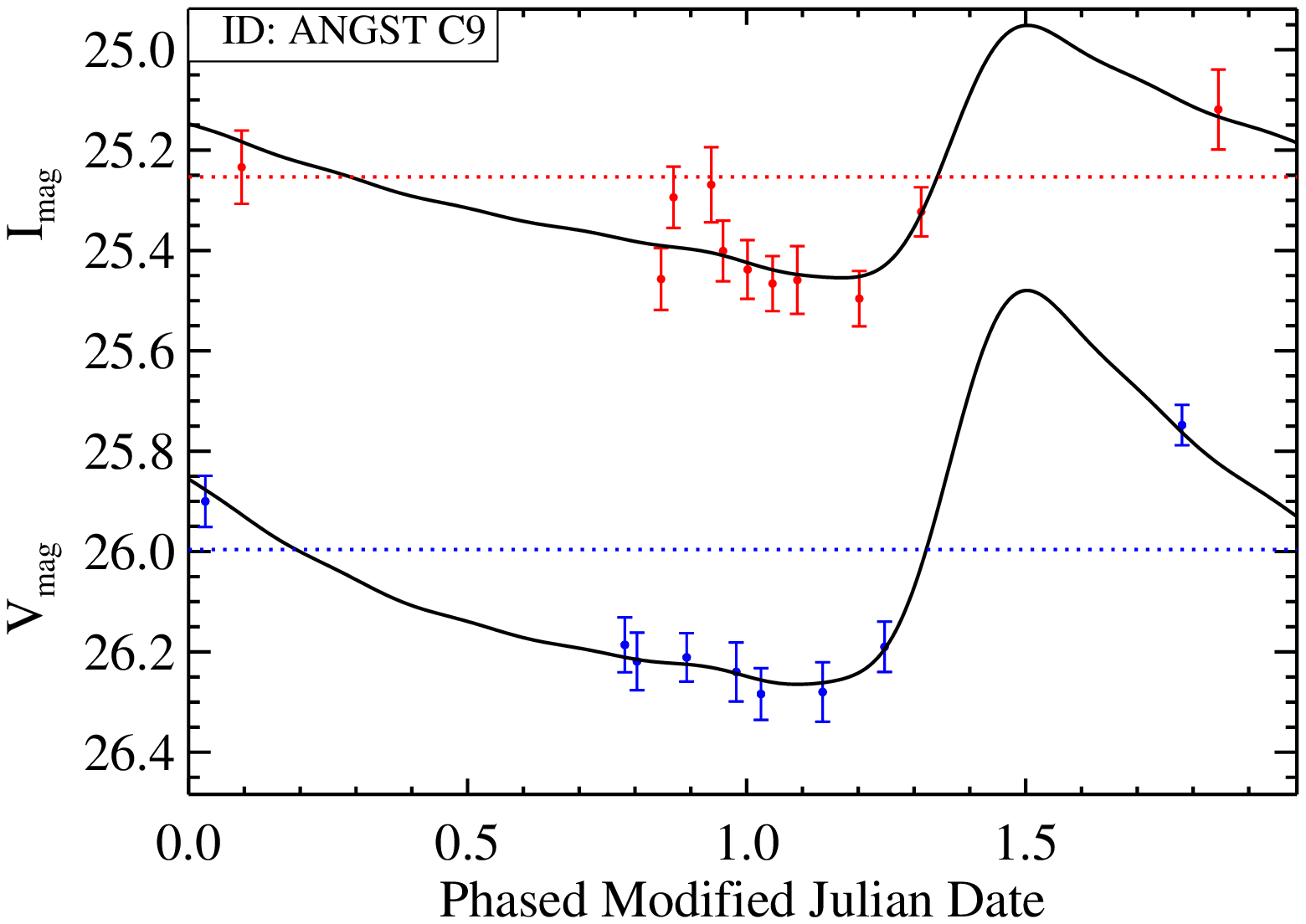} \\  
         \includegraphics[scale=0.3]{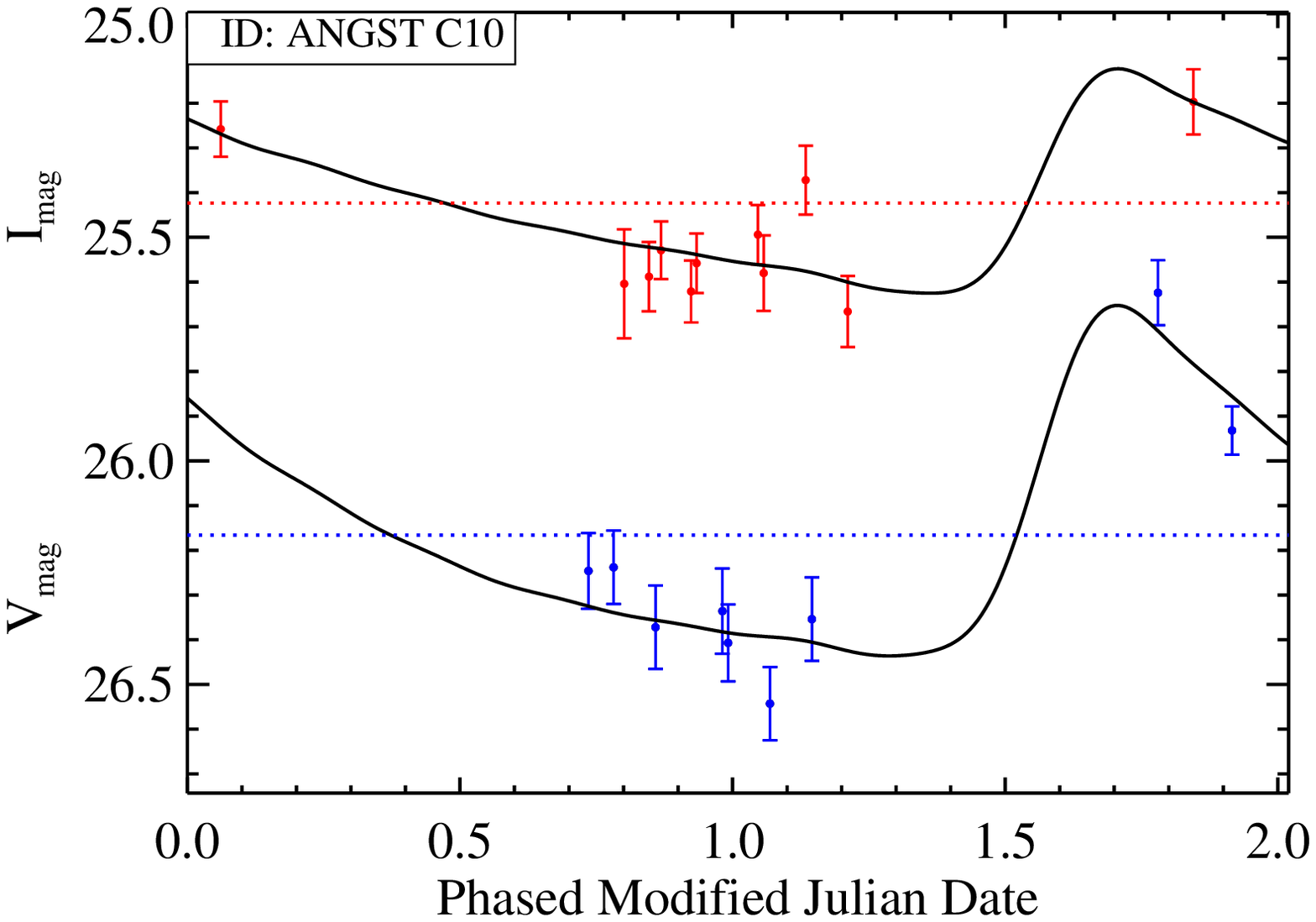} &   
         \includegraphics[scale=0.3]{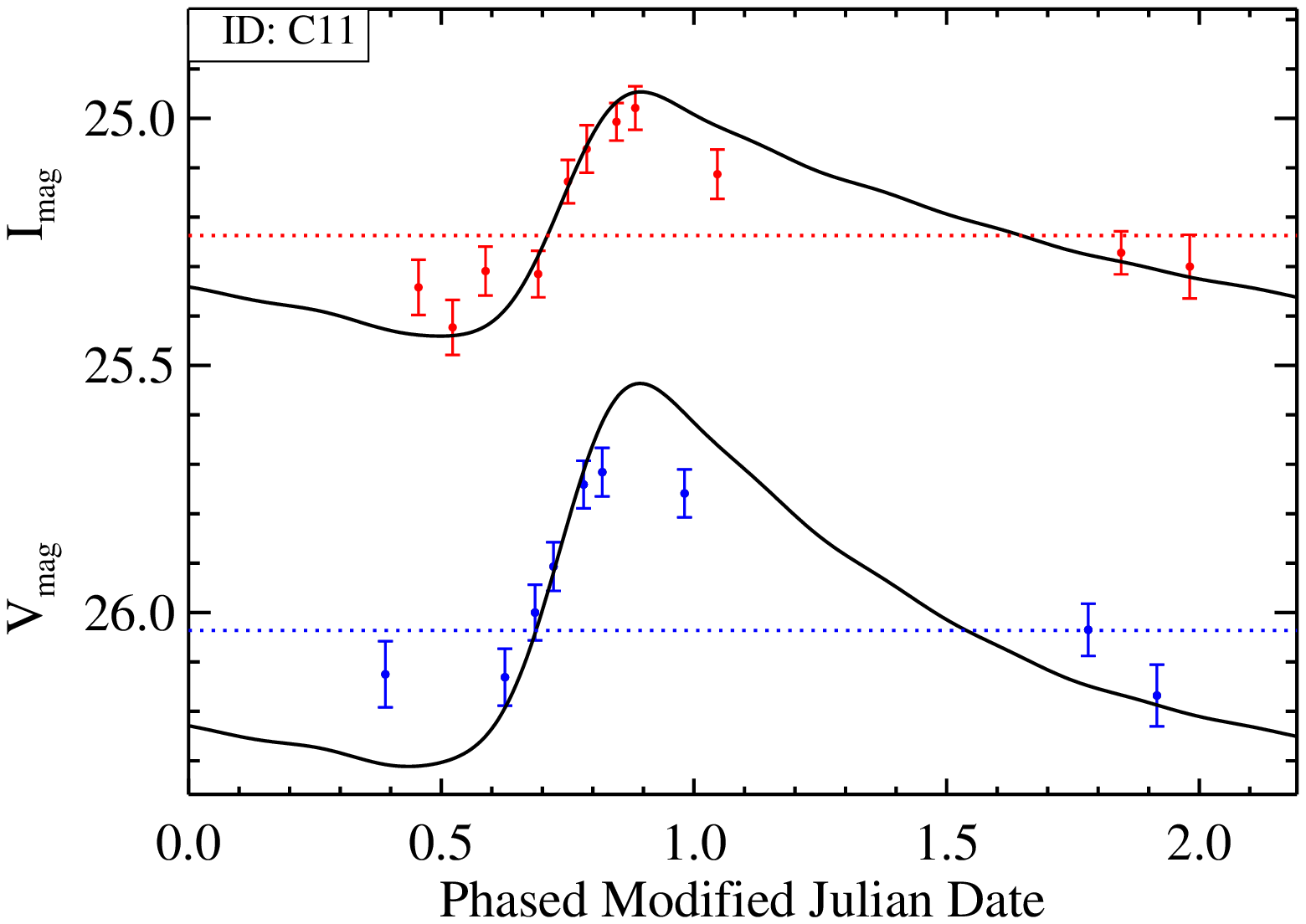}
      \end{array}$
    
 \end{center}
 \caption{M81 fundamental mode Cepheid light curves used for our distance calculation. \label{lc1}}
\end{figure}

\begin{figure}[h]
\plottwo{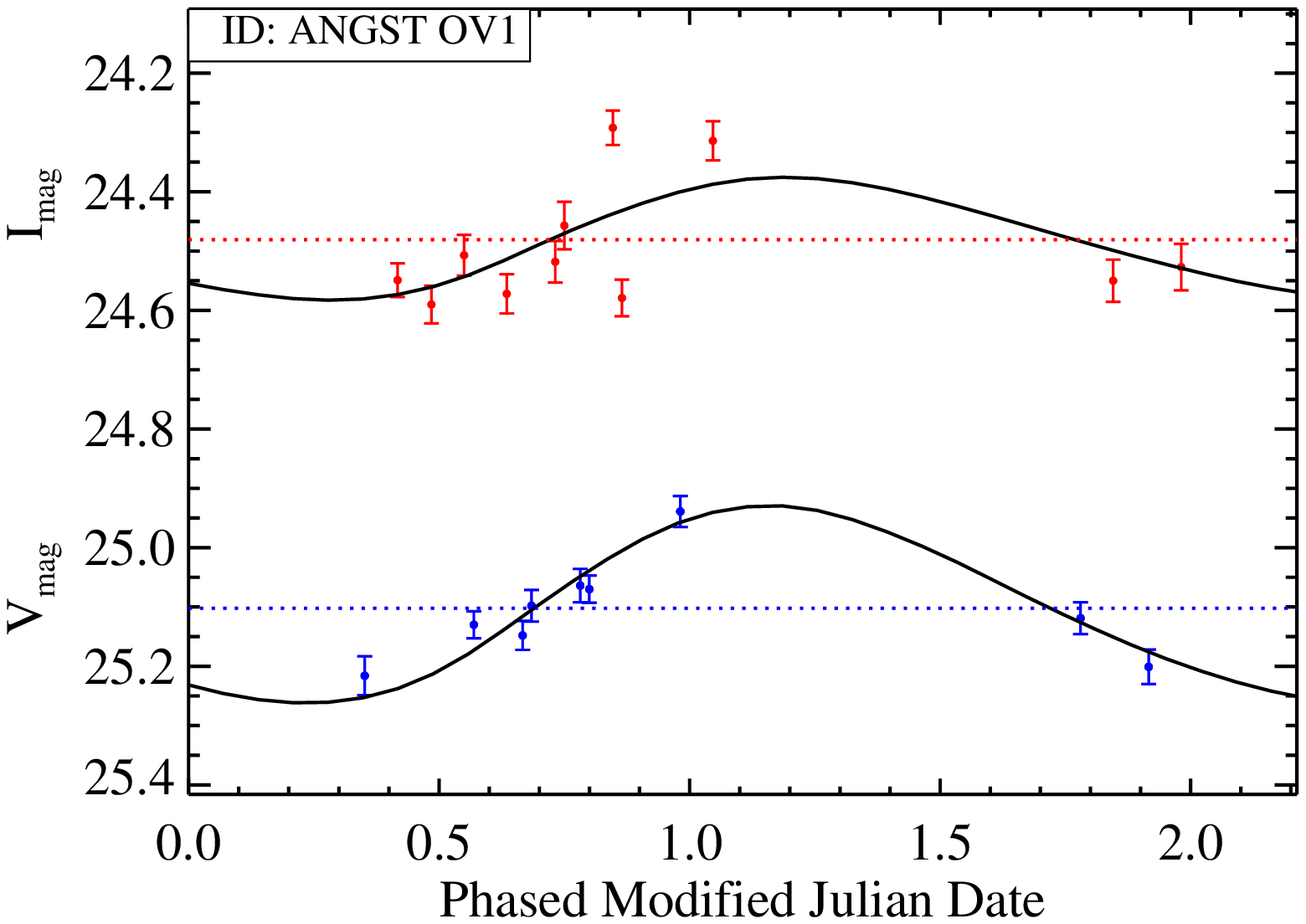}{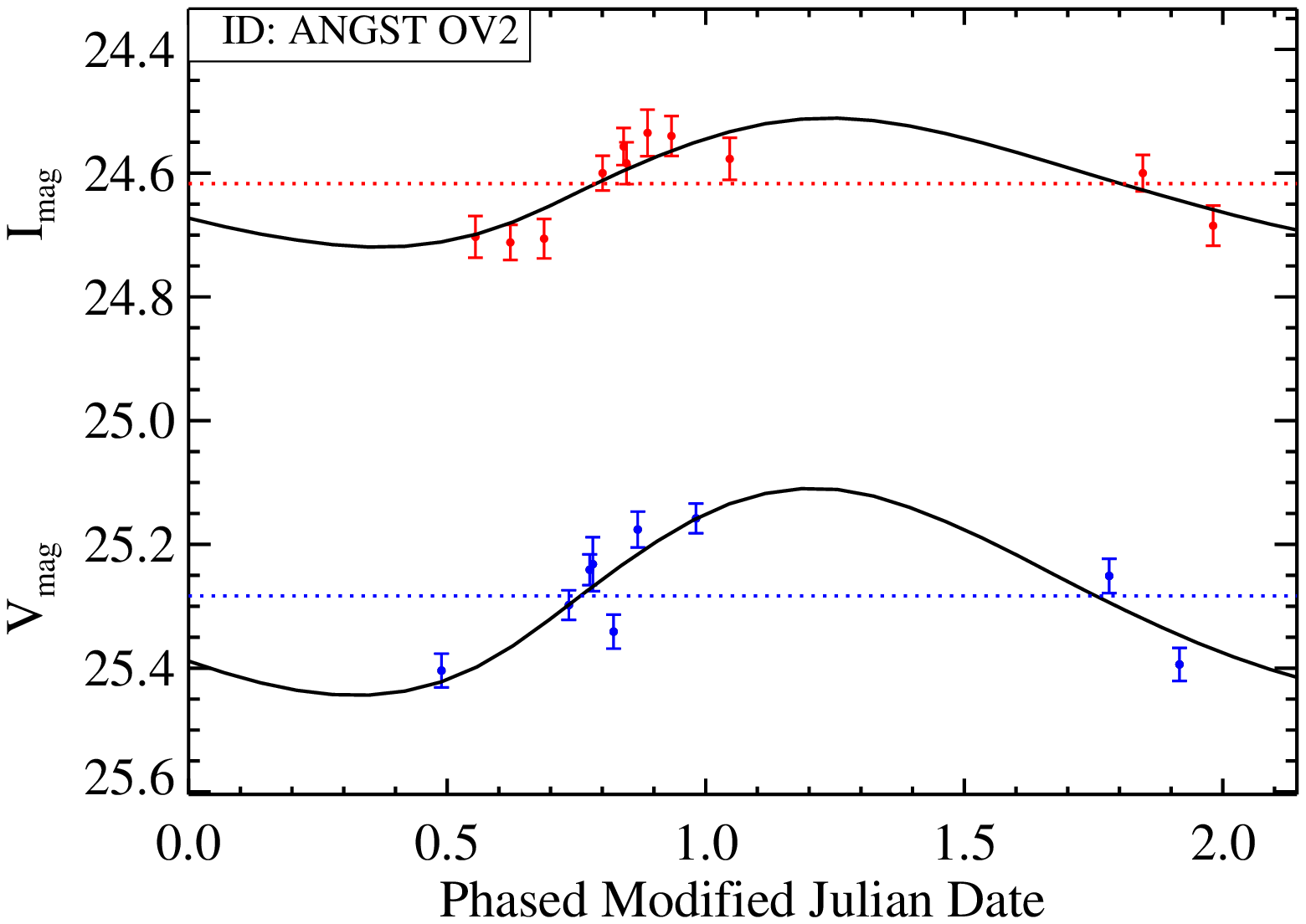}
 \caption{M81 first-overtone Cepheid light curves. \label{lc2}}
\end{figure}


\include{t1}

\end{document}

%% file: t1.tex
\begin{deluxetable*}{ c c c c c  c c c c c}
\tabletypesize{ \scriptsize}
\tablewidth{0pt}
\tablecaption{Photometry for our identified Cepheids \label{full_tab} }
\tablehead{\colhead{Name} & \colhead{MJD-54055} &\colhead{Filter} & \colhead{Mag} & \colhead{error }& \colhead{Name} & \colhead{MJD-54055} &\colhead{Filter} & \colhead{Mag} & \colhead{error }}
\startdata
ANGST C1 & 2.9119 & $V$ & 25.37 &  0.03 & ANGST C8 & 2.9119 & $V$ & 25.46 &  0.04   \\
ANGST C1 & 6.9057 & $V$ & 25.53 &  0.03 & ANGST C8 & 6.9057 & $V$ & 25.37 &  0.04   \\
ANGST C1 & 4.5758 & $V$ & 25.63 &  0.03 & ANGST C8 & 4.5758 & $V$ & 24.87 &  0.08   \\
ANGST C1 & 4.9086 & $V$ & 25.36 &  0.04 & ANGST C8 & 4.9086 & $V$ & 25.02 &  0.30   \\
ANGST C1 & 1.7803 & $V$ & 25.27 &  0.03 & ANGST C8 & 1.7803 & $V$ & 25.15 &  0.07   \\
ANGST C1 & 1.9162 & $V$ & 25.13 &  0.03 & ANGST C8 & 1.9162 & $V$ & 25.33 &  0.03   \\
ANGST C1 & 0.7817 & $V$ & 25.51 &  0.11 & ANGST C8 & 0.7817 & $V$ & 24.97 &  0.08   \\
ANGST C1 & 0.9814 & $V$ & 25.61 &  0.04 & ANGST C8 & 0.9814 & $V$ & 24.97 &  0.08   \\
ANGST C1 & 2.7788 & $V$ & 25.35 &  0.03 & ANGST C8 & 2.7788 & $V$ & 25.40 &  0.04   \\
ANGST C1 & 2.8439 & $I$ & 24.66 &  0.03 & ANGST C8 & 2.8439 & $I$ & 24.65 &  0.04   \\
ANGST C1 & 2.9771 & $I$ & 24.68 &  0.03 & ANGST C8 & 2.9771 & $I$ & 24.61 &  0.04   \\
ANGST C1 & 6.9714 & $I$ & 24.87 &  0.04 & ANGST C8 & 6.9714 & $I$ & 24.61 &  0.04   \\
ANGST C1 & 4.6414 & $I$ & 24.90 &  0.04 & ANGST C8 & 4.6414 & $I$ & 24.18 &  0.07   \\
ANGST C1 & 4.9742 & $I$ & 24.48 &  0.03 & ANGST C8 & 4.9742 & $I$ & 24.23 &  0.07   \\
ANGST C1 & 4.7089 & $I$ & 24.80 &  0.04 & ANGST C8 & 4.7089 & $I$ & 24.27 &  0.07   \\
ANGST C1 & 4.7741 & $I$ & 24.81 &  0.03 & ANGST C8 & 4.7741 & $I$ & 24.35 &  0.03   \\
ANGST C1 & 1.8453 & $I$ & 24.55 &  0.04 & ANGST C8 & 1.8453 & $I$ & 24.45 &  0.03   \\
ANGST C1 & 1.9813 & $I$ & 24.67 &  0.05 & ANGST C8 & 1.9813 & $I$ & 24.62 &  0.05   \\
ANGST C1 & 0.8467 & $I$ & 24.78 &  0.04 & ANGST C8 & 0.8467 & $I$ & 24.40 &  0.03   \\
ANGST C1 & 1.0464 & $I$ & 24.79 &  0.04 & ANGST C8 & 1.0464 & $I$ & 24.47 &  0.04   \\
ANGST C2 & 2.9119 & $V$ & 24.64 &  0.02 & ANGST C9 & 2.9119 & $V$ & 26.28 &  0.05   \\
ANGST C2 & 6.9057 & $V$ & 24.55 &  0.02 & ANGST C9 & 6.9057 & $V$ & 26.19 &  0.05   \\
ANGST C2 & 4.5758 & $V$ & 24.76 &  0.02 & ANGST C9 & 4.5758 & $V$ & 26.22 &  0.06   \\
ANGST C2 & 4.9086 & $V$ & 24.72 &  0.02 & ANGST C9 & 4.9086 & $V$ & 26.28 &  0.06   \\
ANGST C2 & 1.7803 & $V$ & 24.38 &  0.03 & ANGST C9 & 1.7803 & $V$ & 25.75 &  0.04   \\
ANGST C2 & 1.9162 & $V$ & 24.52 &  0.02 & ANGST C9 & 1.9162 & $V$ & 25.90 &  0.05   \\
ANGST C2 & 0.7817 & $V$ & 24.38 &  0.19 & ANGST C9 & 0.7817 & $V$ & 26.19 &  0.05   \\
ANGST C2 & 0.9814 & $V$ & 24.31 &  0.02 & ANGST C9 & 0.9814 & $V$ & 26.24 &  0.06   \\
ANGST C2 & 2.7788 & $V$ & 24.61 &  0.02 & ANGST C9 & 2.7788 & $V$ & 26.21 &  0.05   \\
ANGST C2 & 2.8439 & $I$ & 23.59 &  0.02 & ANGST C9 & 2.8439 & $I$ & 25.40 &  0.06   \\
ANGST C2 & 2.9771 & $I$ & 23.56 &  0.02 & ANGST C9 & 2.9771 & $I$ & 25.46 &  0.07   \\
ANGST C2 & 6.9714 & $I$ & 23.59 &  0.02 & ANGST C9 & 6.9714 & $I$ & 25.32 &  0.05   \\
ANGST C2 & 4.6414 & $I$ & 23.73 &  0.02 & ANGST C9 & 4.6414 & $I$ & 25.29 &  0.06   \\
ANGST C2 & 4.9742 & $I$ & 23.69 &  0.02 & ANGST C9 & 4.9742 & $I$ & 25.50 &  0.06   \\
ANGST C2 & 4.7089 & $I$ & 23.72 &  0.02 & ANGST C9 & 4.7089 & $I$ & 25.27 &  0.07   \\
ANGST C2 & 4.7741 & $I$ & 23.73 &  0.03 & ANGST C9 & 4.7741 & $I$ & 25.44 &  0.06   \\
ANGST C2 & 1.8453 & $I$ & 23.47 &  0.02 & ANGST C9 & 1.8453 & $I$ & 25.12 &  0.08   \\
ANGST C2 & 1.9813 & $I$ & 23.50 &  0.02 & ANGST C9 & 1.9813 & $I$ & 25.23 &  0.07   \\
ANGST C2 & 0.8467 & $I$ & 23.42 &  0.02 & ANGST C9 & 0.8467 & $I$ & 25.46 &  0.06   \\
ANGST C2 & 1.0464 & $I$ & 23.37 &  0.02 & ANGST C9 & 1.0464 & $I$ & 25.47 &  0.05   \\
ANGST C3 & 2.9119 & $V$ & 25.42 &  0.04 & ANGST C10 & 2.9119 & $V$ & 26.41 &  0.09  \\
ANGST C3 & 6.9057 & $V$ & 25.41 &  0.03 & ANGST C10 & 6.9057 & $V$ & 26.35 &  0.09  \\
ANGST C3 & 4.5758 & $V$ & 24.90 &  0.03 & ANGST C10 & 4.5758 & $V$ & 26.25 &  0.08  \\
ANGST C3 & 4.9086 & $V$ & 25.05 &  0.03 & ANGST C10 & 4.9086 & $V$ & 26.54 &  0.08  \\
ANGST C3 & 1.7803 & $V$ & 24.88 &  0.03 & ANGST C10 & 1.7803 & $V$ & 25.62 &  0.07  \\
ANGST C3 & 1.9162 & $V$ & 25.24 &  0.03 & ANGST C10 & 1.9162 & $V$ & 25.93 &  0.05  \\
ANGST C3 & 0.7817 & $V$ & 24.70 &  0.25 & ANGST C10 & 0.7817 & $V$ & 26.24 &  0.08  \\
ANGST C3 & 0.9814 & $V$ & 24.87 &  0.03 & ANGST C10 & 0.9814 & $V$ & 26.34 &  0.10  \\
ANGST C3 & 2.7788 & $V$ & 25.38 &  0.03 & ANGST C10 & 2.7788 & $V$ & 26.37 &  0.09  \\
ANGST C3 & 2.8439 & $I$ & 24.64 &  0.04 & ANGST C10 & 2.8439 & $I$ & 25.62 &  0.07  \\
ANGST C3 & 2.9771 & $I$ & 24.64 &  0.04 & ANGST C10 & 2.9771 & $I$ & 25.58 &  0.08  \\
ANGST C3 & 6.9714 & $I$ & 24.66 &  0.05 & ANGST C10 & 6.9714 & $I$ & 25.67 &  0.08  \\
ANGST C3 & 4.6414 & $I$ & 24.24 &  0.03 & ANGST C10 & 4.6414 & $I$ & 25.60 &  0.12  \\
ANGST C3 & 4.9742 & $I$ & 24.31 &  0.04 & ANGST C10 & 4.9742 & $I$ & 25.37 &  0.08  \\
ANGST C3 & 4.7089 & $I$ & 24.30 &  0.03 & ANGST C10 & 4.7089 & $I$ & 25.53 &  0.06  \\
ANGST C3 & 4.7741 & $I$ & 24.34 &  0.04 & ANGST C10 & 4.7741 & $I$ & 25.56 &  0.07  \\
ANGST C3 & 1.8453 & $I$ & 24.46 &  0.04 & ANGST C10 & 1.8453 & $I$ & 25.20 &  0.07  \\
ANGST C3 & 1.9813 & $I$ & 24.48 &  0.03 & ANGST C10 & 1.9813 & $I$ & 25.26 &  0.06  \\
ANGST C3 & 0.8467 & $I$ & 24.25 &  0.03 & ANGST C10 & 0.8467 & $I$ & 25.59 &  0.08  \\
ANGST C3 & 1.0464 & $I$ & 24.25 &  0.03 & ANGST C10 & 1.0464 & $I$ & 25.49 &  0.07  
\enddata
\end{deluxetable*}

\begin{deluxetable*}{ c c c c c  c c c c c}
\tabletypesize{ \scriptsize}
\tablewidth{0pt}
\tablecaption{Photometry for our identified Cepheids--continued }
\tablehead{\colhead{Name} & \colhead{MJD-54055} &\colhead{Filter} & \colhead{Mag} & \colhead{error }& \colhead{Name} & \colhead{MJD-54055} &\colhead{Filter} & \colhead{Mag} & \colhead{error }}
\startdata
ANGST C4 & 2.9119 & $V$ & 25.77 &  0.04 & ANGST C11 & 2.9119 & $V$ & 25.72 &  0.05  \\	    
ANGST C4 & 6.9057 & $V$ & 25.68 &  0.04 & ANGST C11 & 6.9057 & $V$ & 26.13 &  0.06  \\	    
ANGST C4 & 4.5758 & $V$ & 25.17 &  0.03 & ANGST C11 & 4.5758 & $V$ & 26.12 &  0.07  \\	    
ANGST C4 & 4.9086 & $V$ & 25.32 &  0.03 & ANGST C11 & 4.9086 & $V$ & 25.91 &  0.05  \\	    
ANGST C4 & 1.7803 & $V$ & 25.31 &  0.03 & ANGST C11 & 1.7803 & $V$ & 26.03 &  0.05  \\	    
ANGST C4 & 1.9162 & $V$ & 25.44 &  0.03 & ANGST C11 & 1.9162 & $V$ & 26.17 &  0.06  \\	    
ANGST C4 & 0.7817 & $V$ & 25.08 &  0.03 & ANGST C11 & 0.7817 & $V$ & 25.74 &  0.05  \\	    
ANGST C4 & 0.9814 & $V$ & 25.02 &  0.03 & ANGST C11 & 0.9814 & $V$ & 25.76 &  0.05  \\	    
ANGST C4 & 2.7788 & $V$ & 25.65 &  0.03 & ANGST C11 & 2.7788 & $V$ & 26.00 &  0.06  \\	    
ANGST C4 & 2.8439 & $I$ & 24.83 &  0.05 & ANGST C11 & 2.8439 & $I$ & 25.13 &  0.04  \\	    
ANGST C4 & 2.9771 & $I$ & 24.88 &  0.05 & ANGST C11 & 2.9771 & $I$ & 24.98 &  0.04  \\	    
ANGST C4 & 6.9714 & $I$ & 24.93 &  0.05 & ANGST C11 & 6.9714 & $I$ & 25.32 &  0.05  \\	    
ANGST C4 & 4.6414 & $I$ & 24.59 &  0.03 & ANGST C11 & 4.6414 & $I$ & 25.34 &  0.06  \\	    
ANGST C4 & 4.9742 & $I$ & 24.55 &  0.03 & ANGST C11 & 4.9742 & $I$ & 25.06 &  0.05  \\	    
ANGST C4 & 4.7089 & $I$ & 24.53 &  0.03 & ANGST C11 & 4.7089 & $I$ & 25.42 &  0.06  \\	    
ANGST C4 & 4.7741 & $I$ & 24.47 &  0.03 & ANGST C11 & 4.7741 & $I$ & 25.31 &  0.05  \\	    
ANGST C4 & 1.8453 & $I$ & 24.67 &  0.04 & ANGST C11 & 1.8453 & $I$ & 25.27 &  0.04  \\	    
ANGST C4 & 1.9813 & $I$ & 24.66 &  0.05 & ANGST C11 & 1.9813 & $I$ & 25.30 &  0.06  \\	    
ANGST C4 & 0.8467 & $I$ & 24.48 &  0.04 & ANGST C11 & 0.8467 & $I$ & 25.01 &  0.04  \\	    
ANGST C4 & 1.0464 & $I$ & 24.41 &  0.04 & ANGST C11 & 1.0464 & $I$ & 25.11 &  0.05  \\	    
ANGST C5 & 2.9119 & $V$ & 25.72 &  0.04 & ANGST OV1 & 2.9119 & $V$ & 25.07 &  0.02  \\	    
ANGST C5 & 6.9057 & $V$ & 26.26 &  0.06 & ANGST OV1 & 6.9057 & $V$ & 25.13 &  0.02  \\	    
ANGST C5 & 4.5758 & $V$ & 25.78 &  0.04 & ANGST OV1 & 4.5758 & $V$ & 25.22 &  0.03  \\	    
ANGST C5 & 4.9086 & $V$ & 25.95 &  0.04 & ANGST OV1 & 4.9086 & $V$ & 25.10 &  0.03  \\	    
ANGST C5 & 1.7803 & $V$ & 25.42 &  0.04 & ANGST OV1 & 1.7803 & $V$ & 25.12 &  0.03  \\	    
ANGST C5 & 1.9162 & $V$ & 25.66 &  0.04 & ANGST OV1 & 1.9162 & $V$ & 25.20 &  0.03  \\	    
ANGST C5 & 0.7817 & $V$ & 25.49 &  0.04 & ANGST OV1 & 0.7817 & $V$ & 25.06 &  0.03  \\	    
ANGST C5 & 0.9814 & $V$ & 25.58 &  0.06 & ANGST OV1 & 0.9814 & $V$ & 24.94 &  0.03  \\	    
ANGST C5 & 2.7788 & $V$ & 25.78 &  0.04 & ANGST OV1 & 2.7788 & $V$ & 25.15 &  0.02  \\	    
ANGST C5 & 2.8439 & $I$ & 24.69 &  0.04 & ANGST OV1 & 2.8439 & $I$ & 24.52 &  0.04  \\	    
ANGST C5 & 2.9771 & $I$ & 24.70 &  0.04 & ANGST OV1 & 2.9771 & $I$ & 24.58 &  0.03  \\	    
ANGST C5 & 6.9714 & $I$ & 24.98 &  0.04 & ANGST OV1 & 6.9714 & $I$ & 24.57 &  0.03  \\	    
ANGST C5 & 4.6414 & $I$ & 24.87 &  0.05 & ANGST OV1 & 4.6414 & $I$ & 24.55 &  0.03  \\	    
ANGST C5 & 4.9742 & $I$ & 24.75 &  0.03 & ANGST OV1 & 4.9742 & $I$ & 24.46 &  0.04  \\	    
ANGST C5 & 4.7089 & $I$ & 24.84 &  0.06 & ANGST OV1 & 4.7089 & $I$ & 24.59 &  0.03  \\	    
ANGST C5 & 4.7741 & $I$ & 24.89 &  0.04 & ANGST OV1 & 4.7741 & $I$ & 24.51 &  0.03  \\	    
ANGST C5 & 1.8453 & $I$ & 24.56 &  0.03 & ANGST OV1 & 1.8453 & $I$ & 24.55 &  0.04  \\	    
ANGST C5 & 1.9813 & $I$ & 24.80 &  0.05 & ANGST OV1 & 1.9813 & $I$ & 24.53 &  0.04  \\	    
ANGST C5 & 0.8467 & $I$ & 24.55 &  0.03 & ANGST OV1 & 0.8467 & $I$ & 24.29 &  0.03  \\	    
ANGST C5 & 1.0464 & $I$ & 24.54 &  0.04 & ANGST OV1 & 1.0464 & $I$ & 24.31 &  0.03  \\	    
ANGST C6 & 2.9119 & $V$ & 25.73 &  0.04 & ANGST OV2 & 2.9119 & $V$ & 25.18 &  0.03  \\	    
ANGST C6 & 6.9057 & $V$ & 25.75 &  0.04 & ANGST OV2 & 6.9057 & $V$ & 25.24 &  0.03  \\	    
ANGST C6 & 4.5758 & $V$ & 25.17 &  0.03 & ANGST OV2 & 4.5758 & $V$ & 25.40 &  0.03  \\	    
ANGST C6 & 4.9086 & $V$ & 25.39 &  0.04 & ANGST OV2 & 4.9086 & $V$ & 25.34 &  0.03  \\	    
ANGST C6 & 1.7803 & $V$ & 25.27 &  0.04 & ANGST OV2 & 1.7803 & $V$ & 25.25 &  0.03  \\	    
ANGST C6 & 1.9162 & $V$ & 25.48 &  0.04 & ANGST OV2 & 1.9162 & $V$ & 25.39 &  0.03  \\	    
ANGST C6 & 0.7817 & $V$ & 25.75 &  0.05 & ANGST OV2 & 0.7817 & $V$ & 25.23 &  0.04  \\	    
ANGST C6 & 0.9814 & $V$ & 25.82 &  0.05 & ANGST OV2 & 0.9814 & $V$ & 25.16 &  0.02  \\	    
ANGST C6 & 2.7788 & $V$ & 25.65 &  0.05 & ANGST OV2 & 2.7788 & $V$ & 25.30 &  0.02  \\	    
ANGST C6 & 2.8439 & $I$ & 24.99 &  0.05 & ANGST OV2 & 2.8439 & $I$ & 24.60 &  0.03  \\	    
ANGST C6 & 2.9771 & $I$ & 25.01 &  0.05 & ANGST OV2 & 2.9771 & $I$ & 24.54 &  0.03  \\	    
ANGST C6 & 6.9714 & $I$ & 25.14 &  0.06 & ANGST OV2 & 6.9714 & $I$ & 24.56 &  0.03  \\	    
ANGST C6 & 4.6414 & $I$ & 24.57 &  0.05 & ANGST OV2 & 4.6414 & $I$ & 24.70 &  0.03  \\	    
ANGST C6 & 4.9742 & $I$ & 24.71 &  0.04 & ANGST OV2 & 4.9742 & $I$ & 24.53 &  0.04  \\	    
ANGST C6 & 4.7089 & $I$ & 24.70 &  0.05 & ANGST OV2 & 4.7089 & $I$ & 24.71 &  0.03  \\	    
ANGST C6 & 4.7741 & $I$ & 24.67 &  0.04 & ANGST OV2 & 4.7741 & $I$ & 24.71 &  0.03  \\	    
ANGST C6 & 1.8453 & $I$ & 24.65 &  0.04 & ANGST OV2 & 1.8453 & $I$ & 24.60 &  0.03  \\	    
ANGST C6 & 1.9813 & $I$ & 24.86 &  0.05 & ANGST OV2 & 1.9813 & $I$ & 24.68 &  0.03  \\	    
ANGST C6 & 0.8467 & $I$ & 25.07 &  0.05 & ANGST OV2 & 0.8467 & $I$ & 24.58 &  0.03  \\	    
ANGST C6 & 1.0464 & $I$ & 25.03 &  0.04 & ANGST OV2 & 1.0464 & $I$ & 24.58 &  0.03  \\	    
ANGST C7 & 2.9119 & $V$ & 25.32 &  0.04 & ANGST C7 & 2.8439 & $I$ & 24.92 &  0.05  \\	    
ANGST C7 & 6.9057 & $V$ & 25.81 &  0.05 & ANGST C7 & 2.9771 & $I$ & 24.51 &  0.12  \\	    
ANGST C7 & 4.5758 & $V$ & 25.97 &  0.07 & ANGST C7 & 6.9714 & $I$ & 25.10 &  0.07  \\	    
ANGST C7 & 4.9086 & $V$ & 26.14 &  0.07 & ANGST C7 & 4.6414 & $I$ & 25.06 &  0.06  \\	    
ANGST C7 & 1.7803 & $V$ & 25.81 &  0.06 & ANGST C7 & 4.9742 & $I$ & 25.25 &  0.06  \\	    
ANGST C7 & 1.9162 & $V$ & 26.01 &  0.11 & ANGST C7 & 4.7089 & $I$ & 25.11 &  0.08  \\	    
ANGST C7 & 0.7817 & $V$ & 25.30 &  0.04 & ANGST C7 & 4.7741 & $I$ & 25.21 &  0.07  \\	    
ANGST C7 & 0.9814 & $V$ & 25.42 &  0.08 & ANGST C7 & 1.8453 & $I$ & 24.99 &  0.05  \\	    
ANGST C7 & 2.7788 & $V$ & 25.80 &  0.07 & ANGST C7 & 1.9813 & $I$ & 25.19 &  0.07  \\	    
					 ANGST C7 & 0.8467 & $I$ & 24.78 &  0.05  \\	    
					 ANGST C7 & 1.0464 & $I$ & 24.69 &  0.05          
\enddata
\end{deluxetable*}